\RequirePackage{fix-cm}
\documentclass{svjour3}                     % onecolumn (standard format)
\smartqed  % flush right qed marks, e.g. at end of proof
\usepackage{graphicx}

\usepackage{amsmath} %% for /rightsquigarrow
\usepackage{listings}   
\usepackage{authorcomments}
\usepackage[ruled,vlined,linesnumbered]{algorithm2e}
\usepackage{algorithmicx}
\usepackage{algpseudocode}
\usepackage{subcaption}
\usepackage[export]{adjustbox}
\usepackage{changepage}
\usepackage{natbib}

\newcommand{\code}[1]{\text{\lstinline[basicstyle=\ttfamily, language=JavaScriptColor]~#1~}}

\usepackage{hyperref}

\definecolor{javared}{rgb}{0.6,0,0} % for strings
\definecolor{javagreen}{rgb}{0.25,0.5,0.35} % comments
\definecolor{javapurple}{rgb}{0.5,0,0.35} % keywords
\definecolor{javadocblue}{rgb}{0.25,0.35,0.75} % javadoc

\lstdefinelanguage{JavaScriptColor}{
  keywords={await, async, break, case, catch, const, continue, debugger, default, delete, do, else,
    export, finally, for, function, if, import, in, instanceof, let, of, new, null, return, require, switch, this,
    throw, try, typeof, var, void, while, with, super,
    class, interface, implements, public, private, constructor
    },
  morecomment=[l]{//},
  morecomment=[s]{/*}{*/},
  morestring=[b]',
  morestring=[b]",
  keywordstyle=\color{javapurple}\bfseries,
  identifierstyle=\color{black},
  commentstyle=\color{javagreen}\ttfamily,
  numberstyle=\color{javared}\ttfamily,
  stringstyle=\color{javared}\ttfamily,
  sensitive=false,
  numbers=left,
  stepnumber=1,
  escapeinside={/*\#}{\#*/},
}
\algnewcommand\algorithmicInput{\textbf{Input:}}
\algnewcommand\Input{\item[\algorithmicInput]}

\algnewcommand\algorithmicOutput{\textbf{Output:}}
\algnewcommand\Output{\item[\algorithmicOutput]}

\algnewcommand\algorithmicResult{\textbf{Result:}}
\algnewcommand\Result{\item[\algorithmicResult]}

\algnewcommand\algorithmicforeach{\textbf{for each}}
\algdef{S}[FOR]{ForEach}[1]{\algorithmicforeach\ #1\ \algorithmicdo}
\algdef{SE}[DOWHILE]{Do}{doWhile}{\algorithmicdo}[1]{\algorithmicwhile\ #1}

\algnewcommand\algorithmicpredicate{\textbf{predicate}}
\algdef{SE}[PREDICATE]{Predicate}{EndPredicate}%
   [2]{\algorithmicpredicate\ \textproc{#1}\ifthenelse{\equal{#2}{}}{}{(#2)}}%
   {\algorithmicend\ \algorithmicpredicate}%

\algnewcommand\algorithmicretpred{} % allows us to specify our own return 
\algdef{SE}[RETPRED]{RetPred}{EndRetPred}%
   [2]{\algorithmicretpred\ \textproc{#1}\ifthenelse{\equal{#2}{}}{}{(#2)}}%
   {\algorithmicend\ \algorithmicretpred}%

\newcommand{\tool}{\mbox{\textit{Stubbifier}}\xspace}
\newcommand{\npm}{\mbox{\code{npm}}\xspace}
\newcommand{\nodejs}{\mbox{Node.js}\xspace}

\newcommand{\numTestRuns}{\mbox{10}\xspace}
\newcommand{\numWarmupRuns}{\mbox{two}\xspace}
\newcommand{\numClients}{\mbox{five}\xspace}
\newcommand{\numTotalClients}{\mbox{75}\xspace}
\newcommand{\numProjects}{\mbox{15}\xspace}
\newcommand{\avgStubSizeReduction}{\mbox{56\%}\xspace}
\newcommand{\avgSizeReductionOnBundles}{\mbox{37\%}\xspace}
\newcommand{\avgStubSizeReductionDynamic}{\mbox{31\%}\xspace}
\usepackage{mathptmx}      % use Times fonts if available on your TeX system
\usepackage[symbol]{footmisc}

\usepackage{tcolorbox}
\newenvironment{takeaway}{
\vspace{.5em}
\begin{tcolorbox}[colback=blue!5!white,colframe=blue!5!white,arc=0mm,grow to left by=1.5mm,left=0mm,grow to right by=1.5mm,right=0mm,top=0mm,bottom=0mm]
}
{
\end{tcolorbox}
}
\defcitealias{stubbifierArtifact}{Turcotte and Arteca, et al., 2021} % <-- new

\begin{document}

\title{Stubbifier: Debloating Dynamic Server-Side JavaScript Applications
\thanks{
This research was supported in part by Office of Naval Research (ONR) grants N00014-17-1-2945 and N00014-21-1-2491, and by National Science Foundation grant CCF-1907727.
E. Arteca and A. Turcotte are supported in part by the Natural Sciences and
Engineering Research Council of Canada.}
%Grants or other notes}
}
% \subtitle{Do you have a subtitle?\\ If so, write it here}

\author{Alexi Turcotte\footnote{\label{note}These authors contributed equally to the work.}  
\and Ellen Arteca\footref{note}
\and Ashish Mishra \and Saba Alimadadi \and Frank Tip
}

\authorrunning{A. Turcotte and E. Arteca et al.} % if too long for running head

\institute{A.Turcotte, E.Arteca, F.Tip \at
              Northeastern University,
              Boston, MA, USA,\\ 
              \email{\{turcotte.al, arteca.e, f.tip\}@northeastern.edu}           
           \and
           A.Mishra \at
              Purdue University, West Lafayette, IN, USA\\
              \email{mishr115@purdue.edu}
            \and 
           S.Alimadadi \at 
             Simon Fraser University, Vancouver, Canada\\
             \email{saba@sfu.ca}
}

\date{}

\maketitle

\renewcommand{\thefootnote}{\arabic{footnote}}

\vspace*{-1cm}
\begin{abstract}

JavaScript is an increasingly popular language for server-side development, thanks in part to the Node.js runtime environment and its vast ecosystem of modules.
With the Node.js package manager {\tt npm}, users are able to easily include external modules as dependencies in their projects.
However, {\tt npm} installs modules with {\it all} of their functionality, even if only a fraction is needed, which causes an undue increase in code size.
Eliminating this unused functionality from distributions is desirable, but the sound analysis required to find unused code is difficult due to JavaScript's extreme dynamicity.

We present a fully automatic technique that identifies unused code by constructing static or dynamic call graphs from the application's tests, and replacing code deemed unreachable with either file- or function-level {\it stubs}. 
If a stub is called, it will fetch and execute the original code on-demand, thus relaxing the requirement that the call graph be sound. 
The technique also provides an optional \textit{guarded execution mode} to guard application against injection vulnerabilities in untested code that resulted from stub expansion.

This technique is implemented in an open source tool called \tool, which supports the ECMA\-Script 2019 standard.
In an empirical evaluation on \numProjects Node.js applications and \numTotalClients clients of these applications, \tool reduced application size by \avgStubSizeReduction on average while incurring only minor performance overhead.
The evaluation also shows that \tool's guarded execution mode is capable of preventing several known injection vulnerabilities that are manifested in stubbed-out code.
Finally, \tool can work alongside {\it bundlers}, popular JavaScript tools for bundling an application with its dependencies. For the considered subject applications, 
we measured an average size reduction of \avgSizeReductionOnBundles in bundled distributions.

\end{abstract}

\section{Introduction}

% Intro to JavaScript
JavaScript is one of the most popular programming languages, and has been the lingua franca of client-side web development for years~\citep{la:githubtoplanguages,stackoverflow2020}.
More recently, platforms such as \nodejs\citep{nodejs}
have made it possible to use JavaScript outside of the browser.
\nodejs provides a light-weight, fast, and scalable platform for writing network-based applications, enabling web developers to use the same language for both front- and back-end development.
As a result, server-side JavaScript development has experienced an exponential growth in recent years.

% NPM and code bloat
This has given rise to a flourishing ecosystem of libraries, known as Node modules, that are freely available and widely used.
The \npm\citep{npm}
package-management system in particular has fostered higher developer productivity and increased code reuse by unburdening the programmers from many routine development tasks.
As such, a typical Node module $m$ can directly and indirectly rely on myriad other modules.
While an essential attribute of this ecosystem, in practice, $m$ typically uses only a small fraction of the functionality of its dependencies, while still encompassing all of their code.
In turn, clients of $m$ inherit the unused functionality of $m$ and its dependencies, as well as that of its own dependencies.
The problem of accumulating code that in practice is never invoked is known as code ``bloat''.

% Debloating
While eliminating code bloat is desirable, ``debloating'' \nodejs applications is challenging since it is nearly impossible to perform sound static analysis on JavaScript due to the high dynamism of the language.
% Eliminating code bloat is desirable as reducing the size of an application can improve its performance, and decrease the amount of development and maintenance efforts required from developers. \SA{citation?} % find studies on size of apps and their performance + maintenance/debugging efforts
% Code bloat can quickly arise in \nodejs application development, where it is universal practice for open-source and commercial applications to rely on modules available on \npm
% Unfortunately, ``debloating'' \nodejs applications is challenging since it is nearly impossible to perform sound static analysis on JavaScript due to the high dynamism of the language.
%
Despite the popularity of \nodejs development and the severity of this issue, there is currently no technique available that can significantly debloat a modern \nodejs application and still preserve its complete original behavior.
% Bundlers.
JavaScript \textit{bundlers} are applications whose primary focus is to create self-contained application distributions, and these bundlers perform an optimization known as ``tree-shaking''~\citep{TreeShakingMDN}
on imported external modules, by removing
modules or functions that are unreachable in an application's import graph. Unfortunately, the size reduction achieved by bundlers is limited by the all-or-nothing nature of their code minimization technique: 
code that the bundler removes must \textit{never} be called, else the bundled application will crash. 
% and these bundlers perform rudimentary debloating that falls short of meaningfully reducing application size.

% Previous research
Previous work on debloating in the context of other languages has focused on the use of static analysis to determine
unreachable code~\citep{AgesenUngar:94,VisualAgeSmalltalk:97,VisualWorks:95,DBLP:journals/toplas/TipSLES02}.
% The highly dynamic features of \nodejs applications make it difficult for traditional static analysis techniques to carry out debloating effectively, as the inevitable unsoundness of the analysis may lead to removal of code that is deemed unused by mistake.
In many existing techniques, the application stops executing when trying to invoke code that has been removed by the debloating algorithm and deviates from the intended behavior of the original application.
Despite more recent advances for analyzing client-side web applications~\citep{DBLP:conf/sigsoft/LivshitsK08,andreasen2014,DBLP:conf/sigsoft/JensenMM11,DBLP:journals/pacmpl/LiTMS18,DBLP:conf/sigsoft/LiTMS18,DBLP:conf/ecoop/SridharanDCST12}, the development of a static analysis for \nodejs that is simultaneously sound, precise, and scalable remains beyond the current state of the art.

% Proposed technique
% In this paper, we present a novel technique for reducing the size of \nodejs applications while preserving the original behavior, and implement this technique in a tool called \tool.
% Our approach is inspired by the \textsc{Doloto} tool \citep{DBLP:conf/sigsoft/LivshitsK08}, and comprises a combination of unsound static and dynamic analysis with a technique known as {\it code splitting} pioneered by \textsc{Doloto}.
% Our technique expands on \textsc{Doloto} in its support for modern JavaScript \citep{ECMAScript2019}, in its addition of file-level stubs, and importantly in its fully-automatic nature; whereas \textsc{Doloto} required traces of users interacting with the subject application do establish code reachability, 
% our approach debloats projects automatically by using the project's tests to determine what code is reachable.
% For a given application, \tool uses its tests as a model of its usage and infers a call graph, determines which functions and files are untested, and instead of {\it removing} this code outright, {\it replaces} it with stub versions that fetch and execute the original code dynamically.
% As stubbed code represents untested code, \tool also allows users to opt-in to a {\it guarded execution mode}, where stubbified code is modified to intercept calls to functions such as \code{eval} and \code{exec} that can execute arbitrary user code.
% The reduced \nodejs application is then ready to be deployed, exercised, and reused by other applications.

This paper presents a practical technique for reducing the size of \nodejs applications while preserving their original behavior. 
Core application functionality, as well as the extent to which an application uses its dependencies, is inferred automatically from dynamic or static call graphs constructed from the application's own test suites (which can be comprehensive, end-to-end test suites).
Untested, unreachable code is replaced by {\it stub} versions in a technique known as {\it code splitting}, pioneered by the \textsc{Doloto} tool \citep{DBLP:conf/sigsoft/LivshitsK08}.
If a function or file stub is executed, it will dynamically fetch and execute the original code so as to preserve application functionality.
The technique has been implemented in a tool called \tool, which improves on \textsc{Doloto} by:
  (i)   supporting all features of modern JavaScript \cite{ECMAScript2019}, including classes, promises, async/await, generators, and modules,
  (ii)  introducing file-level stubs in addition to function-level stubs (so as to achieve additional debloating by stubbing all code in files where no code is used, instead of stubbing each of the functions in these files individually), and
  (iii) providing an (optional) {\it guarded execution mode}, where stubbed-out code is automatically instrumented to intercept calls to 
        functions such as \code{eval} and \code{exec} that may introduce injection vulnerabilities.
Most importantly, (iv) \tool is \textit{fully automatic} by relying on static analysis or dynamic analysis of the application's test suite 
to identify code that is likely to be unreachable, whereas \textsc{Doloto} required %\TODO{manually gathered?} 
traces of users interacting with the subject application to establish core application functionality.

\tool was evaluated on \numProjects of the most popular Node.js applications, using \numClients clients for each subject application to evaluate
how much code is loaded dynamically. This evaluation found that \tool achieves significant size reductions (\avgStubSizeReduction on average), that
the number of stubs expanded during the execution of client applications is relatively small, and that minimal performance overhead is incurred.
Further, experiments with \tool's guarded execution mode confirmed that it is capable of preventing known injection vulnerabilities.
Finally, we confirmed experimentally that, when used in conjunction with the popular \texttt{Rollup} bundler, \tool achieves significant additional 
size reductions on previously bundled applications (\avgSizeReductionOnBundles on average). 
% \TODO{please review the above para}

In summary, this paper makes the following contributions:
\begin{itemize} % contriubutions

  \item % technique
	A fully automated technique for reducing the size of \nodejs applications while preserving their original behavior, 
	based on a combination of static or dynamic analysis and code splitting.
%	Through a combination of dynamic and static analysis along with code splitting, our approach allows the reduced version to dynamically load and execute \textit{debloated} code, and ensures that the original behavior of a reduced application is preserved. 
		
  \item % implementation
	The implementation of this technique in a tool called \tool that supports modern JavaScript \cite{ECMAScript2019}. 
	\tool is publicly available as an open-source tool\footnote{
	  See \href{https://github.com/emarteca/stubbifier}{\textcolor{blue}{https://github.com/emarteca/stubbifier}}.
	}, and a self-contained code artifact including reproducible experiments is also available on Zenodo \citepalias{stubbifierArtifact}.  
	
  \item % evaluation
	An empirical evaluation of \tool on \numProjects open source \nodejs applications and \numTotalClients clients of these subject applications (\numClients clients per subject),
	showing that 
	  \tool reduces the size of \nodejs applications by \avgStubSizeReduction on average while incurring only minor performance overhead. 
    The evaluation also shows that \tool's guarded execution mode is capable of preventing several known injection vulnerabilities that
    are manifested in stubbed-out code.
	
\end{itemize}

\section{Background and Motivation}

The \npm ecosystem includes more than 1.7 million modules%
\footnote{
  See \url{http://www.modulecounts.com/}.
} that provide a wealth of convenient features. By importing these libraries and reusing their functionality, programmers can focus their efforts on
features that are unique to their application.  
However, this convenience does not come without its price: importing modules can cause projects to become excessively large due to the transitive importing of other
projects that they depend on. 
In practice, it is often the case that a module only uses a small subset of the functions in the transitive closure of its dependencies.

To illustrate this, consider the example of a popular node application {\tt css-loader} \citep{css-loader-npm}, a utility package for loading, parsing, and transforming {\tt CSS} files and further supporting applications designed to use {\tt CSS}.
{\tt css-loader} is one of the most popular modules on \npm, with nearly 15 million weekly downloads, and it is imported by over 15,000 modules. 

{\tt css-loader} has 13 third-party production dependencies%
\footnote{
  Many npm modules rely on additional development dependencies (sometimes referred to as ``devDependencies'') that are needed only
  for development purposes, e.g., for running tests. These dependencies are typically not installed by clients.
} (i.e., modules upon which its functionality depends).
% \TODO{it is not clear what is meant by a "production dependency and a "core-module" dependency, so I think you need to define these terms.
    %   It is also confusing that the term "development dependency is used in section 4. Should use consistent terminology}
The stand-alone {\tt css-loader} module contains only 16 files comprising 110KB. 
However, installing {\tt css-loader} with direct and transitive production dependencies creates a package with 1299 files and total code size of 2764KB. 
This is a $>$81x and $>$25x increase in number of files and code size respectively.

To determine what part of the resulting installation constitutes application bloat, we examined {\tt css-loader} to determine which functions and files are reachable from the application's test suite.
Using a simple static analysis that traces function calls  to build a list of unreachable functions and files, 209 files were found to be potentially unreachable, and 6 unreachable functions  were identified in otherwise reachable files.
Given the extreme dynamicity of JavaScript, sound static analysis is not possible
(see, e.g., \cite{DBLP:conf/ecoop/RichardsHBV11,DBLP:conf/ecoop/SridharanDCST12,DBLP:conf/issta/JensenJM12,DBLP:conf/oopsla/MadsenTL15,DBLP:journals/pacmpl/SteinNCM19}).
Since in practice all static analyses for JavaScript suffer from unsoundness, some of the functions and files that they identify as being unreachable may indeed be reachable.
Nevertheless, if one {\it could} devise a technique to remove all of this code, the application's size would be reduced by 80\%.

Consider {\tt semver}~\citep{semver-npm}, a package that {\tt css-loader} depends on.
%{\tt semver} is the ``semantic versioner for \npm''~\citep{semver-npm}.
Only \emph{two} functions from {\tt semver} are used in {\tt css-loader}:
The {\tt satisfies} function is imported specifically as part of the primary {\tt css-loader} functionality, and the {\tt inc} function is used once as a helper in the {\tt css-loader} tests.
% satisfies: \href{https://github.com/webpack-contrib/css-loader/blob/master/src/index.js#L9}{\textcolor{blue}{here}}.
% inc: \href{https://github.com/webpack-contrib/css-loader/blob/master/test/helpers/ast-loader.js#L6}{\textcolor{blue}{here}}
Given that, it seems wasteful to include the {\it entire} {\tt semver} code in {\tt css-loader}, and indeed \tool identifies 27 of {\tt semver}'s files and six {\tt semver} functions inside of {\tt css-loader} to be potentially unreachable.
The six unreachable functions are in the file that exports the {\tt inc} function.
After debloating, the code in the {\tt semver} package is reduced from 57KB to 35KB, a 38\% size reduction.

As we will discuss in Section~\ref{sec:evaluation}, our \tool tool reduces the size of {\tt css-loader} by 80\%, from 2.8MB to 0.6MB. 
{\tt css-loader} has approximately 14.8 million weekly downloads, so an 80\% size reduction would  translate to a reduction in weekly data transfer from 41.4TB to 8.8TB. 

The next sections will present our debloating technique and its evaluation.  

%\TODO{from intro}
%To illustrate the impact of this size reduction, consider the module \href{https://www.npmjs.com/package/css-loader}{\textcolor{blue}{{\tt css-loader}}}, which has 14.8 million weekly downloads: \tool achieves an 80\% size reduction (from 2.8MB to 0.6MB) in this application, which would translate to a reduction in weekly data transfer from 41.44TB to 8.8TB.

% \TODO{note: content about bundlers was integrated with section 3.4 and content about Doloto was integrated with related work}   

\section{Approach}

 The debloating technique presented in this paper involves several key steps:
   constructing call graphs,
   introducing stubs,
   optionally introducing security checks to safeguard against injection vulnerabilities in stubbed-out code, and
   integration with bundlers.
The remainder of this section will discuss each of these steps in detail.
 
\subsection{Call Graph Construction}
\label{sec:call-graph}

In principle, any call graph can be used to determine which files and functions should be replaced with stubs. The soundness 
and precision of the call graph  will impact the size of the initial distribution and the amount of code that needs to be loaded dynamically.

The implementation of \tool includes mechanisms for constructing a static or dynamic call graph.
In each case, \tool uses the test suite of the input application as the entry point for the analysis, and so the call graph represents the 
\emph{tested} code.
Any function that is not in the call graph is deemed unreachable and untested and will be replaced with a (\emph{function-level} or \emph{file-level}) stub.
% -- this is what will be stubbed.
Both analyses are configured to consider depended-upon modules (in the \code{node_modules} directory), though note that development dependencies are excluded as they are typically not packaged and shipped with the subject application. 
% \TODO{see earlier comment that we need to use clearly defined and consistent terminology when discussing dependences}

Below, we provide some further detail on the specific static and dynamic call graph construction techniques that \tool supports.

\paragraph{Dynamic Call Graphs.}
To compute \emph{dynamic} call graphs, code coverage is determined using Istanbul's command line tool {\tt nyc} \citep{istanbulNPM}, %\TODO{use citations for references to URLs}, 
that computes statement, line, branch, and function coverage for Node.js applications.
By default, {\tt nyc} ignores a project's dependencies, but \tool automatically generates a configuration file that specifies that coverage of \textit{non-development}, production dependencies should be computed.
\tool then runs {\tt nyc} on the application's tests, to determine which functions and files are invoked during testing (and by exclusion, which were not invoked).

\paragraph{Static Call Graphs.}
To compute the \emph{static} call graphs, we developed an analysis using CodeQL~\citep{QLpaper},  GitHub's declarative language for static 
analysis, using  its extensive libraries for writing static analyzers~\citep{qlrepo}. In particular, CodeQL's dataflow library contains functionality 
for tracking calls through local module imports, and we implemented an extension to recognize modules in a project's {\tt node\_modules} directory, 
and extended CodeQL's libraries to track data flow through these modules. Then, a call graph construction algorithm was implemented on top of this
analysis, which uses the application's tests as entry points for the analysis.

\subsection{Introducing Stubs}
\label{sec:generating-stubs}

After constructing a call graph, \tool creates lists of unreachable functions and files.
Here, \textit{unreachable files} are those in which \emph{none} of the functions are reachable, and \textit{unreachable functions} are 
those functions that are not reachable but that are in a file where at least one other function \emph{is} reachable.

Next, \tool parses the application's source code, including any dependencies, and replaces unreachable functions and files with 
\emph{stubs} via transformations on the program's Abstract Syntax Tree (AST).  Note that \tool does not replace functions or files 
with stubs if they are shorter than the stubs that would replace them.

\paragraph{File Stubs.}
Each unreachable file is replaced with a \textit{file stub}.
The code in this stub implements Algorithm~\ref{alg:file-stub}, which depicts the general logic for file stub expansion. % the code that needs to be executed if a stubbed file is expanded.
\begin{algorithm}
 perform all imports\;
 
 let $\mathit{file_o}$ := fetchOriginalFileCode()\;
 let $\mathit{file_e}$ := {\bf eval}($\mathit{file_o}$)\;
 
 replace this file with $\mathit{file_o}$\;
 
 perform all exports\;
 \caption{ExpandFileStub\label{IR}}
 \label{alg:file-stub}
\end{algorithm}

\noindent
At a high level, file stub expansion amounts to:
  (i) performing all imports that were in the original code (line 1), 
  (ii) fetching the original code and evaluating it (lines 2-3), 
  (iii) replacing the contents of the stubbed file with the original file (line 4), and finally 
  (iv) performing necessary exports (line 5).
More specifically, in files that rely on the CommonJS mechanisms (i.e., \code{require} for importing and \code{module.exports} for exporting), 
simply storing the original code elsewhere and \code{eval}-ing it as needed suffices, as these mechanisms can be used anywhere in a source file. %operate dynamically.
% The addition of the ECMAScript Module System (ESM) in 2015~\citep{ecma1999262} provides an added complication to our approach.
However, the ECMAScript Module System (ESM)~\citep{ECMAScriptSpecification}'s static \code{import}/\code{export} constructs cannot be executed in an 
\code{eval} (see section 15.2 of \cite{ECMAScript2019}), so all \code{import} and \code{export} statements are hoisted out of the 
original code and into the stub. The original code is then transformed to properly produce the values of the exports.
To illustrate, consider the  example in Figure~\ref{fig:file-stub-example-pre}.
\begin{figure}[htbp]
\begin{center}
\begin{lstlisting}
// file.js before stubbification
export function foo() { /* ... */ }
import { A };
function bar() { /* ... */ }
export default bar;


// file.js after stubbification
import { A };
exportObj = eval(stubs.getCodeForFile("file.js"));/*#\label{line:file-eval}#*/

let foo_UID = exportObj["foo"];
export {foo_UID as foo};/*#\label{line:namedexportstub}#*/
export default exportObj["default"]/*#\label{line:defaultexpstub}#*/
\end{lstlisting}
\caption{File before and after stubbification.}
\label{fig:file-stub-example-pre}
\end{center}
\end{figure}

In Figure~\ref{fig:file-stub-example-pre}, we see \code{import} and \code{export} statements interspersed through the file before stubbification.
In the lower part of the figure, we see that the file stub generated by \tool contains all \code{import} statements as-is, and \code{export} statements are modified (lines~\ref{line:namedexportstub} and~\ref{line:defaultexpstub}) to get their values from the dynamically executed code (i.e., from \code{exportObj}, line~\ref{line:file-eval}).
% To generate the file after stubbification, \tool finds all \code{import} statements, removes them from the code, and places them at the top of the stub.
% Additionally, \tool finds all of the file's \code{export}s, and generates new export statements like those on lines~\ref{line:namedexportstub} and~\ref{line:defaultexpstub}.
% of Figure~\ref{fig:file-stub-example-post}.

% To allow this exporting, % the return value of \code{eval} must specify exactly what should be exported, thus \tool also transforms the code that would be passed to \code{eval} if the stub were expanded.
To allow this exporting, the original code from Figure~\ref{fig:file-stub-example-pre} is modified to construct an object containing all of the original exports.
This constructed object is the last statement that will be executed when the code is passed to \code{eval}, and is therefore the return value of \code{eval}.
This is illustrated in Figure~\ref{fig:modif-code-ex}.
\begin{figure}[htbp]
\begin{center}
\begin{lstlisting}
function foo() { /* ... */ }
function bar() { /* ... */ }

{  foo: foo,/*#\label{line:fooExpo}#*/
   default: bar };
\end{lstlisting}
\caption{Modified original code with ES6 imports and exports (this is what would be \code{eval}'d).}
\label{fig:modif-code-ex}
\end{center}
\end{figure}

Here, we see that the \code{export} was removed from the definition of \code{foo}, and that \code{foo} was added to an object on line~\ref{line:fooExpo}, which also includes an entry for \code{bar}, the default export.
The last statement in an \code{eval}-ed code block is implicitly returned---here, that is an object containing the exports---allowing the stub to retrieve the exported values (as in line~\ref{line:file-eval} of Figure \ref{fig:file-stub-example-pre}).

\paragraph{Function Stubs.}

% \TODO{this section needs to explain how our function stubs are different from the ones created by Doloto. If they are the same, we should acknowledge that}

Functions deemed unreachable are replaced with \textit{ function stubs}. These stubs implement Algorithm~\ref{alg:func-stub}, which depicts the 
general logic for dynamically loading and executing code upon stub expansion.
\begin{algorithm}
 \KwData{$\mathbf{args}$: function arguments}
 \KwData{$\mathit{uid}$: unique ID for this function stub}
 
 \uIf{$uid$ cached}{
  let $\mathit{fun_{str}}$ := code cached at $\mathit{uid}$\;
 }\Else{
  let $\mathit{fun_{str}}$ := fetch original function code\;
 }

 let $\mathit{fun_e}$ := {\bf eval}($\mathit{fun_{str}}$)\;
 copy function properties to $\mathit{fun_e}$\;
 
 \uIf{can replace function definition}{
  replace stub with $\mathit{fun_e}$\;
 }\Else{
  cache $\mathit{fun_{str}}$\;
 }
 
 call $\mathit{fun_e}$ with $\mathbf{args}$\;
 return result\;
 
 \caption{ExpandFunctionStub\label{IR}}
 \label{alg:func-stub}
\end{algorithm}
\noindent
When a stub is expanded, it first fetches the code, either by retrieving it from a cache, fetching it from a server, or otherwise retrieving it from storage.
Either way, the code is evaluated into a function, and function properties are copied from the stub version to the newly created function object.
If possible, \tool will replace the stub with the freshly evaluated original function (the conditions where this is or is not possible are discussed below). 
% \TODO{Clarify when this is possible}
If not, the code is cached, and then the function is executed.

\tool's caching strategy differs from \textsc{Doloto}'s~\citep{DBLP:conf/sigsoft/LivshitsK08}: where \textsc{Doloto} caches function objects, \tool caches the code, and we discuss the reasoning behind this shortly.

A concrete example of a function stub can be found in Figure~\ref{fig:fun-stub-example}, where we show the stub for \code{getValidHeaders} from the {\tt node-blend}~\citep{node-blend} project. % the function \code{getValidHeaders} from the {\tt node-blend} project which was transformed by our tool.
\begin{figure}[htbp]
\begin{center}
\begin{lstlisting}
function getValidHeaders(headers) {
  let toExec = eval(stubs.getCode("UID_for_LOC"));/*#\label{line:stubGetCodeCall}#*/
  stubs.cpFunProps(getValidHeaders, toExec);/*#\label{line:copyFuncProps}#*/
  getValidHeaders = toExec;/*#\label{line:reassignFunction}#*/
  return toExec.apply(this, arguments);/*#\label{line:applyArgs}#*/
}
\end{lstlisting}
\caption{Example of a stubbed function.}
\label{fig:fun-stub-example}
\end{center}
\end{figure}

First, note that \tool outfits each file with a global \code{stubs} object containing the code cache and functionality to fetch code. 
% \TODO{if there is a stubs object for each file, it is not "global". Or is there a single stubs object containing the stubs for all files?}
% We implement a function stub ``manager'' object, that provides an interface that function stub expansions can use for fetching the code to be loaded.
% \tool instantiates one of these objects as \code{stubs} for each file that contains function stubs -- 
We see a call to \code{stub.getCode("UID_for_LOC")} on line~\ref{line:stubGetCodeCall}, which fetches the \textit{original function definition} (found through \code{"UID_for_LOC"}, a unique ID for the function that \tool generates from the code location when the stub is created).
That code is then passed to \code{eval}, which will return a function object containing the original code.
Line~\ref{line:copyFuncProps} copies any function properties from \code{getValidHeaders} to the fresh function\footnote{
Recall that in JavaScript functions are objects, and can have properties assigned dynamically.
}.
Finally, line~\ref{line:reassignFunction} redefines the \code{getValidHeaders} with the expanded stub, and line~\ref{line:applyArgs} calls the function with its original arguments\footnote{
\code{apply} calls its receiver as a function, binding its first argument to \code{this} inside the function, and passing the other arguments as function arguments.
\code{arguments} is a metavariable available inside functions that refers to its arguments.
}.
Since \code{getValidHeaders} has reassigned itself on line~\ref{line:reassignFunction}, any subsequent calls to this function  will call the expanded stub, with no need to re-\code{eval} the code.

The above discussion covered the general approach for introducing function stubs. However, several types of functions require special treatment,
as will be discussed next.

\paragraph{\bf Anonymous Functions.}
% \subsubsection{Functions with no ID}
In JavaScript it is possible to create a function without a name, an idiom that is commonly seen when functions are passed as callback arguments to higher-order functions.
In these cases, the function cannot reassign itself as is done on line~\ref{line:reassignFunction} in the above example (since it has no name to refer to itself by), 
so the loaded code is cached, and future stub expansions \code{eval} the cached code.
For example, Figure~\ref{fig:stub-anon} displays the \code{getValidHeaders} stub that we would create if this function did not have a name.
% \begin{lstlisting}
% function(headers) {
%     let toExecString = stubs.getStub("UID_for_LOC");/*#\label{line:stubCacheAcc}#*/
%     if (! toExecString) {
%         toExecString = stubs.getCode("UID_for_LOC");
%         stubs.setStub("UID_for_LOC", toExecString);
%     }
%     let toExec = eval(toExecString);
%     toExec = stubs.cpFunProps(this, toExec);
%     return toExec.apply(this, arguments);
% }
% \end{lstlisting}
\begin{figure}
    \begin{lstlisting}
function(headers) {
    let toExecString = stubs.getStub("UID_for_LOC");/*#\label{line:stubCacheAcc}#*/
    if (! toExecString) {
        toExecString = stubs.getCode("UID_for_LOC");
        stubs.setStub("UID_for_LOC", toExecString);
    }
    let toExec = eval(toExecString);
    toExec = stubs.cpFunProps(this, toExec);
    return toExec.apply(this, arguments);
}
\end{lstlisting}
    \caption{Example of stubbed anonymous function.}
    \label{fig:stub-anon}
\end{figure}
Here, rather than immediately passing the code loaded with \code{stubs.getCode("UID_for_LOC")} to \code{eval}, the \code{stubs} cache is accessed on line~\ref{line:stubCacheAcc}.
Code is only loaded on a cache miss, in which case the loaded code is immediately cached.
% Code loading only occurs on a cache miss, in which case it is immediately cached so as to only fetch the code once.

One might wonder why the function stub expansion caches the loaded code, evaluating it every time the stub is invoked, rather than caching the expanded function object.
This is necessary because, in JavaScript, functions  are \textit{closures} that {\it close} variables from surrounding scopes directly into the object.
Therefore,  generating a stub for a function that is nested inside another would include the {\it function arguments} of the latter in its closure.
If we were to cache this object, any subsequent call to the function would refer to the values of function arguments when the stub was first expanded, which may lead to incorrect program behavior.
Thus, we have to \code{eval} every time.
Note that this problem does not arise for functions with a name, as the function reassigning itself does not store a closure.
% \TODO{reworded the above discussion slightly -- please review carefully. Did Doloto handle this case correctly?}

\textsc{Doloto} cached function closures, which is problematic for the reasons discussed above; we conjecture that the authors did not 
evaluate their tool on code where this issue would arise.
% either because only outer functions are replaced with stubs, or they were wrong and just didn't run into the bug we ran into.

\paragraph{\bf Class and Object Methods.}
% \subsubsection{Class and Object Methods}
When replacing object or class methods with stubs, an issue arises that relates to references to \code{this}.
In functions outside a class or object, \code{this} refers to the function object itself, while in a class/object, \code{this} refers to the \emph{object instance} on which the function was invoked.
These class methods need to be referenced in a different way to allow for function property copying and reassignment.
% Thus, we need another way of referencing the function from inside itself, so that function properties can properly be copied, and so that the function can reassign itself (if it has an ID).

Fortunately, class and object methods can be accessed as \emph{properties} of \code{this}, and so if \code{getValidHeaders} were a method in a class, the following replacements would be made:
\begin{lstlisting}
// outside a class/object
stubs.cpFunProps(getValidHeaders, toExec);
getValidHeaders = toExec;

// inside a class/object
stubs.cpFunProps(this.getValidHeaders, toExec);
this.getValidHeaders = toExec;
\end{lstlisting}

For class/object methods with no ID, we generate a dynamic property access on \code{this} to reassign the function object as at code generation time, we know the key corresponding to nameless object properties.
Specifically, this means that instead of \code{this.functionName} we use \code{this[\$\{generate(key)\}\$]}, where \code{\$\{generate(key)\}\$} is a string generated at parsing runtime, to reference the function as a dynamic property access on \code{this}.
% \AT{Fix this paragraph.}

Classes and objects can also have \emph{getter} and \emph{setter} methods, as is illustrated in the example below:
\begin{lstlisting}
class A {
    get propName() { console.log("getter"); }
    set propName() { console.log("setter"); }
}
let x = new A();
x.a; // prints "getter"
x.a = 5; // prints "setter"
\end{lstlisting}
Getter and setter stubs are generated with special reassignment code. % , as they are not amenable to traditional method reassignment.
Dynamically accessing and defining a getter for some property \code{"p"} is done using \code{this.__lookupGetter__("p")} and   \code{this.__defineGetter__("p")} respectively (and similarly for setters); these calls are used in place of direct accesses as properties of \code{this} in the stub.

\paragraph{\bf Arrow Functions.}
% \subsubsection{Arrow functions}
Arrow functions were introduced in ECMAScript 2015, and provide a more concise syntax for functions.
When creating stubs for arrow functions, we run into an issue as the metavariable \code{arguments} cannot be used to reference the function arguments.
To get around this, we make use of the \emph{rest parameter}~\citep{rest-parameter-docs}, also introduced with ES6.
By replacing the original function parameters with a rest parameter, we have essentially recreated the functionality of \code{arguments}.
For example, if \code{getValidHeaders} were an arrow function, it would be written as:
\begin{lstlisting}
let getValidHeaders = (headers) => { /* elided function body */ }
\end{lstlisting}
and its stub would resemble:
\begin{lstlisting}
let getValidHeaders = (...args_UID) => {
    // only change the last line of the stub
    getValidHeaders.apply(this, args_UID);
}
\end{lstlisting}

\paragraph{\bf Unstubbable Functions.}
% \subsubsection{Functions that cannot be stubbed}
\tool does not transform generators, as \code{yield} cannot be present inside of an \code{eval}, nor does it transform constructors.
Constructors necessitate that \code{super} be called before any use of the \code{this} keyword.
Generating constructor stubs would require a more sophisticated analysis of constructor code, and as sound static analysis of JavaScript is still very challenging, we decided against stubbifying them altogether.
%
% \tool is also unable to generate stubs for generators, as \code{yield} cannot be present inside of an \code{eval}.
%
We do not consider this to be a big issue, 
% as these types of functions are 
% \TODO{Some of this does not make sense to me---if a class is used, then its constructor is generally called. Perhaps you mean to say that it is
% rare for a class to be used without using its constructor? Please clarify}
as these types of functions are fairly rare; we only encountered a few instances of unreachable constructors or generators in our evaluation.

\paragraph{{\bf Manually specifying functions \emph{not} to stub.}} \label{dontStubMeSection}
Users may be interested in specifying some functions that should never be replaced with stubs, regardless of their classification in the generated call graph.
To accommodate this, we added functionality to allow users to manually flag a function so it will be ignored by \tool.

% To avoid a function ever being stubbed, users the following line of code as the first line in the function in question: \code{eval("STUBBIFIER_DONT_STUB_ME");}.

% For example, given a function
% \begin{lstlisting}
% function dontStubMe() { 
%   console.log("hello");
%   // ...
% }
% \end{lstlisting}
% To specify this function should never be stubbed, rewrite it as:
% \begin{lstlisting}
% function dontStubMe() { 
%   eval("STUBBIFIER_DONT_STUB_ME");
%   console.log("hello");
%   // ...
% }
% \end{lstlisting}
\subsection{Guarded Execution Mode}

Since \tool builds the input call graph using the application's tests, the stubbed code is also the \emph{untested} code.
Dynamically loading and executing this code could pose a security risk, as it may include injection vulnerabilities that
were not encountered during testing.

To address such concerns,  \tool includes an option to detect calls to a pre-specified list of ``dangerous'' functions in expanded code.
This is achieved by intercepting all function calls and checking whether or not the function is (perhaps an alias of) one of these dangerous functions.
In our current implementation, the list of these functions consists of: \code{eval}, \code{process.exec}, and \code{child_process.\{fork, exec, execSync, spawn\}}, common functions that enable the execution of arbitrary code. % \code{child_process.fork}, \code{child_process.exec},\\ \code{child_process.execSync}, and \code{child_process.spawn}, common functions that enable the execution of arbitrary code.
% list consists of common functions that execute code dynamically or create new processes: \code{eval}, \code{exec}, \code{fork}, and \code{spawn}.
It is trivial to include other functions to this list, so users can customize what functions they want guard against.
We include an example of the code with guards in the supplementary material.

These checks can be configured to generate a warning, or exit the application if a dangerous function is about to be called.
% In our evaluation, we have the check generate a warning, so that we can easily track the number of these functions that are called.
This transformation is run on the original code so that, when a stub is expanded, the loaded code includes guards.
% \TODO{But we only introduce these checks in code that is dynamically loaded, right?  Right now, it sounds like we're doing this instrumentation
%   everywhere, so please clarify}

Since these functions could be aliased, we must wrap \emph{every} function call with these checks.
As such, the size of the loaded code (i.e., the expanded stubs) is increased dramatically. 
% \TODO{can we say "the size of the code that stubs expand to is increased dramatically"?}
The guards also incur more runtime overhead, as will be discuss in Section~\ref{sec:evaluation}. 
\subsection{Bundler Integration}
\label{sec:impl:bundler}

Many Java\-Script projects use {\it bundlers} such as {\tt webpack} \citep{webpackNPM} and {\tt Rollup} \citep{rollupNPM} to package 
an application along with all the modules that it depends on into a single-file distribution that includes all required functionality. 
Such a bundle can be included in another application using \code{require} or \code{import}, so that users do not need to go through 
additional installation steps. 

Bundlers perform a limited form of code debloating known as ``tree-shaking'', which identifies functions and classes
that are unused based on a static analysis of the import relationships between modules. If the project relies on \code{require} statements 
to import external functionality, the required files are simply included in the bundle in their entirety; if the project relies on the ECMAScript 
Module System, parts of an imported module can be removed if they are not referenced in the importing module.  The size reductions that
can be achieved using tree-shaking can be significant, but they are still limited by the fact that soundness is required, because
the removal of code that is used could cause a bundled application to crash.

The use of \tool in combination with a bundler requires a few additional steps in the previously discussed transformation pipeline.
First, bundling must always happen \emph{before} applying \tool, as bundlers perform their own code transformations. For example, 
when merging all the application code into a single file, 
bundlers often refactor the code so as to avoid variable name conflicts, repeated imports, etc.
If \tool were run on the application before bundling, the bundler would only perform its analysis on the code that is not replaced with stubs, since the 
code to be loaded dynamically is just stored as plain text. 
As a result, expanding a stub would result in code that does not match, e.g., the changed variable names in the bundle, which is likely to result in errors. 

To prevent such issues, \tool should be applied to an application \textit{after} it has been processed by a bundler. 
One minor obstacle here is that \tool uses an application's tests as the entry points for call graph construction, and
tests are nearly always based on the original project source code, and not on a bundle. 
To address this, \tool determines a mapping of the functions to be stubbed from their positions in the original code to their 
positions in the bundle.  Then, it constructs call graphs from tests as discussed before, and it consults the mapping to determine where
stubs should be introduced in the bundle. 

The evaluation presented in Section~\ref{sec:evaluation} will examine how much additional code size reduction can be achieved by \tool on 
applications after they have been bundled using {\tt Rollup}.

\section{Evaluation and Discussion}
\label{sec:evaluation}

This section presents an evaluation of \tool that aims to answer the following research questions:
\begin{itemize}
	\item {\bf RQ1.} How much does \tool reduce application size, and which type of call graphs (static or dynamic) is more effective for
	                reducing application size?
	\item {\bf RQ2.} How much code is dynamically loaded due to stub expansion?
	\item {\bf RQ3.} How much overhead is incurred due to stub expansion? 
	\item {\bf RQ4.} How much time does \tool need to transform applications?
	\item {\bf RQ5.} How much run-time overhead is incurred by guarded execution mode and can it detect security vulnerabilities?
	\item {\bf RQ6.} How much does \tool reduce the size of applications that have been bundled using \texttt{Rollup}?
\end{itemize}

\noindent
% We answer these research questions through a series of experiments on popular JavaScript applications.

%
%
%
%
\subsection{Experimental Setup and Methodology}

To evaluate \tool, we selected 15 projects from the most popular projects published by \npm; from a list of projects sorted in descending order by number of weekly downloads, we attempted to install, build, and run project test suites (as \tool uses the test suite to generate call graphs).
If a project satisfied all these criteria, we then randomly selected from its {\it dependents}, or {\it clients}, and attempted to install, build, and run their tests; if the project had five such clients, it was selected.

Table~\ref{table:projDescTable} lists the projects used for the evaluation, as well as some relevant metrics.
The first row reads: the project \code{memfs} has 18k lines of code (LOC) in the analyzed files%
\footnote{
  The metrics in the table reflect the project's own source code (excluding tests), and all its (transitive) production dependencies, but
  excluding \code{devDependencies}.
.
}, and there are 133 files analyzed (Num files).
The \code{memfs} test suite has 284 tests (Tests), one production dependency, and its analyzed code comprises 146 KB (Size). 

We have created \href{https://doi.org/10.5281/zenodo.5599914}{\textcolor{blue}{a code artifact}} \citepalias{stubbifierArtifact} to accompany this paper: the artifact includes each project cloned at the version on which we ran the evaluation, the experimental infrastructure used to conduct said evaluation, as well as the full source code of \tool.

\begin{table*}
 
  \centering

  % \subfloat[][]{
  \begin{subtable}{\textwidth}
      \centering
      \footnotesize
      % \adjustbox{max width=0.97\textwidth}{
      \begin{tabular}{ l | c || c | c | c | c | c } 
      \multicolumn{6}{c}{} \\
      \multicolumn{6}{c}{} \\
      \textbf{Project (citation)} & \textbf{Commit} & \textbf{LOC} & \textbf{\# files} & \textbf{Tests} & \textbf{Prod deps} & \textbf{Size (KB)} \\
      \hline\hline
      \href{https://github.com/streamich/memfs}{\textcolor{blue}{\texttt{memfs}} (\cite{memfs})} & {\tt a9d2242} & 18k & 133 & 284 & 1 & 146 \\
      \href{https://github.com/bdistin/fs-nextra}{\textcolor{blue}{\texttt{fs-nextra}} (\cite{fsnextra})} & {\tt 6565c81} & 11k & 184 & 138 & 0 & 52 \\
      \href{https://github.com/expressjs/body-parser}{\textcolor{blue}{\texttt{body-parser}} (\cite{bodyparser})} & {\tt 480b1cf} & 20k & 210 & 231 & 21 & 364 \\
      \href{https://github.com/tj/commander.js}{\textcolor{blue}{\texttt{commander}} (\cite{commander})} & {\tt 327a3dd} & 13k & 177 & 351 & 0 & 70 \\
      \href{https://github.com/webpack/memory-fs}{\textcolor{blue}{\texttt{memory-fs}} (\cite{memoryfs})} & {\tt 3daa18e} & 14k & 167 & 44 & 11 & 120 \\
      \href{https://github.com/isaacs/node-glob}{\textcolor{blue}{\texttt{glob}} (\cite{glob})} & {\tt f5a57d3} & 13k & 175 & 1706 & 10 & 86 \\
      \href{https://github.com/reduxjs/redux}{\textcolor{blue}{\texttt{redux}} (\cite{redux})} & {\tt b5d07e0} & 105k & 4491 & 82 & 2 & 267 \\
      \href{https://github.com/webpack-contrib/css-loader}{\textcolor{blue}{\texttt{css-loader}} (\cite{cssloader})} & {\tt dcce860} & 71k & 1299 & 430 & 36 & 2764 \\
      \href{https://github.com/kriskowal/q}{\textcolor{blue}{\texttt{q}} (\cite{q})} & {\tt 6bc7f52} & 16k & 135 & 243 & 0 & 281 \\
      \href{https://github.com/pillarjs/send}{\textcolor{blue}{\tt send} (\cite{send})} & {\tt de073ed} & 14k & 157 & 152 & 17 & 97\\
      \href{https://github.com/expressjs/serve-favicon}{\textcolor{blue}{\tt serve-favicon} (\cite{servefavicon})} & {\tt 15fe5e3} & 10k & 121 & 30 & 5 & 20\\
      \href{https://github.com/expressjs/morgan}{\textcolor{blue}{\tt morgan} (\cite{morgan})} & {\tt 19a6aa5} & 14k & 159 & 81 & 8 & 55\\
      \href{https://github.com/expressjs/serve-static}{\textcolor{blue}{\tt serve-static} (\cite{servestatic})} & {\tt 94feedb} & 13k & 160 & 90 & 19 & 106 \\
      \href{https://github.com/facebook/prop-types}{\textcolor{blue}{\tt prop-types} (\cite{proptypes})} & {\tt d62a775} & 15k & 152 & 287 & 4 & 106 \\
      \href{https://github.com/expressjs/compression}{\textcolor{blue}{\tt compression} (\cite{compression})} & {\tt 3fea81d} & 13k & 149 & 38 & 11 & 66 \\
      \end{tabular}
      % \end{adjustbox}
      \vspace*{1mm}
      % }
      \caption{Summary of projects used for evaluation}\label{table:projDescTable}
  \end{subtable}
%   \hfill
  
  % \subfloat[][]{
  \begin{subtable}{\textwidth}
    %  \vspace{10pt}
     \centering
     \footnotesize
      % \adjustbox{max width=0.97\textwidth}{
      \begin{tabular}{ l | c | c | c | c }
    %   \multicolumn{5}{c}{{\bf Static CG}} \\
    %   \hline
      {\bf Project} & \textbf{Size (KB)} & {\bf Reduction \%} & {\bf Expanded (KB)} & {\bf Red after exp (\%)} \\
      \hline\hline
      \href{https://github.com/streamich/memfs}{\textcolor{blue}{\texttt{memfs}}} & 19 & 87\% & [19, 138] & [87\%, 5\%] \\
      \href{https://github.com/bdistin/fs-nextra}{\textcolor{blue}{\texttt{fs-nextra}}} & 31 & 39\% & [31, 45] & [39\%, 14\%]  \\
      \href{https://github.com/expressjs/body-parser}{\textcolor{blue}{\texttt{body-parser}}} & 65 & 82\% & [211, 297] & [42\%, 18\%]   \\
      \href{https://github.com/tj/commander.js}{\textcolor{blue}{\texttt{commander}}} & 68 & 2\% & [68, 68] & [2\%, 2\%]   \\
      \href{https://github.com/webpack/memory-fs}{\textcolor{blue}{\texttt{memory-fs}}} & 41 & 66\% & [41, 87] & [66\%, 27\%] \\
      \href{https://github.com/isaacs/node-glob}{\textcolor{blue}{\texttt{glob}}} & 61 & 28\% & [70, 80]  & [18\%, 7\%] \\
      \href{https://github.com/reduxjs/redux}{\textcolor{blue}{\texttt{redux}}} & 201 & 25\% & [221, 221]  & [17\%, 17\%] \\
      \href{https://github.com/webpack-contrib/css-loader}{\textcolor{blue}{\texttt{css-loader}}} & 559 & 80\% & [559, 895] & [80\%, 68\%]  \\
      \href{https://github.com/kriskowal/q}{\textcolor{blue}{\texttt{q}}} & 37 & 87\% & [37, 100] & [87\%, 64\%] \\
      \href{https://github.com/pillarjs/send}{\textcolor{blue}{\tt send}} & 59 & 39\% & [59, 92] & [39\%, 5\%] \\
      \href{https://github.com/expressjs/serve-favicon}{\textcolor{blue}{\tt serve-favicon}} & 15 & 24\% & [15, 18] & [24\%, 8\%]   \\
      \href{https://github.com/expressjs/morgan}{\textcolor{blue}{\tt morgan}} & 25 & 55\% & [41, 45] & [25\%, 20\%]  \\
      \href{https://github.com/expressjs/serve-static}{\textcolor{blue}{\tt serve-static}} & 38 & 64\% & [38, 83] & [64\%, 21\%]  \\
      \href{https://github.com/facebook/prop-types}{\textcolor{blue}{\tt prop-types}} & 18 & 83\% & [56, 56] & [48\%, 48\%]  \\
      \href{https://github.com/expressjs/compression}{\textcolor{blue}{\tt compression}} & 24 & 63\% & [24, 24] & [63\%, 63\%]
      \end{tabular}
      % \end{adjustbox}
      \vspace*{1mm}
      % }
      \caption{Size of projects stubbified with static CG}\label{table:SizeRQStaticTable}
  \end{subtable}

  \begin{subtable}{\textwidth}
    %  \vspace{10pt}
     \centering
     \footnotesize
      % \adjustbox{max width=0.97\textwidth}{
      \begin{tabular}{ l | c | c | c | c }
    %   \multicolumn{5}{c}{{\bf Dynamic CG}} \\
    %   \hline
       {\bf Project} & \textbf{Size (KB)} & {\bf Reduction \%} & {\bf Expanded (KB)} & {\bf Red after exp (\%)} \\
      \hline\hline
      \href{https://github.com/streamich/memfs}{\textcolor{blue}{\texttt{memfs}}} & 17 & 89\% & [17, 136] & [89\%, 7\%]\\
      \href{https://github.com/bdistin/fs-nextra}{\textcolor{blue}{\texttt{fs-nextra}}} & 47 & 10\% & [47, 47] & [10\%, 10\%] \\
      \href{https://github.com/expressjs/body-parser}{\textcolor{blue}{\texttt{body-parser}}} & 173 & 53\% & [180, 253]  & [51\%, 31\%] \\
      \href{https://github.com/tj/commander.js}{\textcolor{blue}{\texttt{commander}}} & 59 & 16\% & [59, 59]  & [16\%, 16\%] \\
      \href{https://github.com/webpack/memory-fs}{\textcolor{blue}{\texttt{memory-fs}}} & 100 & 17\% & [100, 117] & [17\%, 3\%]\\
      \href{https://github.com/isaacs/node-glob}{\textcolor{blue}{\texttt{glob}}} & 84 & 4\% & [91, 91]  & [-6\%, -6\%]\\
      \href{https://github.com/reduxjs/redux}{\textcolor{blue}{\texttt{redux}}} & 189 & 29\% & [209, 209] & [22\%, 22\%] \\
     \href{https://github.com/webpack-contrib/css-loader}{\textcolor{blue}{\texttt{css-loader}}} &  584 & 79\% & [584, 1372]  & [79\%, 50\%]\\
      \href{https://github.com/kriskowal/q}{\textcolor{blue}{\texttt{q}}} & 206 & 27\% & [206, 209] & [27\%, 26\%]\\
      \href{https://github.com/pillarjs/send}{\textcolor{blue}{\tt send}} & 89 & 8\% & [89, 93] & [8\%, 5\%]\\
      \href{https://github.com/expressjs/serve-favicon}{\textcolor{blue}{\tt serve-favicon}} & 19 & 3\% & [19,19 ]  & [3\%, 3\%]\\
      \href{https://github.com/expressjs/morgan}{\textcolor{blue}{\tt morgan}} & 49 & 13\% & [52, 52] & [7\%, 7\%] \\
      \href{https://github.com/expressjs/serve-static}{\textcolor{blue}{\tt serve-static}} & 98 & 7\% & [98, 102] & [7\%, 4\%] \\
      \href{https://github.com/facebook/prop-types}{\textcolor{blue}{\tt prop-types}} & 16 & 85\% & [53, 53]  & [50\%, 50\%]\\
      \href{https://github.com/expressjs/compression}{\textcolor{blue}{\tt compression}} & 46 & 29\% & [46, 46] & [29\%, 29\%] \\
      \end{tabular}
      % \end{adjustbox}
      \vspace*{1mm}
      % }
      \caption{Size of projects stubbified with dynamic CG}\label{table:SizeRQDynamicTable}
  \end{subtable}
  \caption{}
\end{table*}

\paragraph{Selecting subject applications.}

Each subject application was processed twice with \tool, once using static call graphs and once using dynamic call graphs.
In each case, files and functions deemed unreachable were replaced with stubs. To address {\bf RQ4}, the time required for the
entire process was measured. For {\bf RQ1}, the size of the application before and after introducing stubs was compared.
We compute the size of source code (excluding tests), \textit{including} production dependencies and \textit{excluding} development dependencies.

To address {\bf RQ2} and {\bf RQ3}, we selected \numClients clients of each subject package from its list of dependents that is published on \npm.
These clients were essentially selected randomly, but we excluded clients without tests or with failing tests.  We also confirmed that the 
dependency is actually used in the client: there are some projects that list a package as a dependency but no longer use in the source code, 
and we excluded these. Finally, we exclude clients that require the use of older versions of \nodejs.

\paragraph{Conducting performance measurements.}

To determine the performance overhead caused by stub expansion, we compared the runtime of each of these clients' tests when using the 
stubbed and original subject application.  When running the test suite with the stubbed application, we also tracked the total size and 
number of stub expansions to determine \emph{how much} code is loaded dynamically. %  when the stubbed application is used.
In our evaluation, stubs were loaded from local storage on the machine running the evaluation.

To mitigate noise and bias caused by caching, all test suites were executed \numTestRuns times after \numWarmupRuns test runs 
before the timed experiments; the reported results are the average of these \numTestRuns test runs.
Furthermore, since some of the tests generate files in {\tt /tmp}, this directory is cleared between every test suite run.
% 
% \TODO{eliminated some text here that seemed to provide unnecessary detail} 

Finally, to mitigate versioning errors, we run our experiments on a client using the same version of the dependency as the one that we transform.
Specifically, we do the following when testing a client:
\begin{itemize}
    \item {\tt npm} or {\tt yarn install} in the root of the client project.
    \item Replace the dependency in question in the client's {\tt node\_modules} with a \emph{symbolic link} to the source code of the dependency that we will transform.
    \item Run the client's tests.
    \item Transform the dependency. 
    The symbolic link means the client needs no change to use the stubbed version of the dependency. % on the client's side: the dependency has been transformed.
    \item Rerun the client's tests, now with the stubbed version of the dependency. % , now with the transformed version of the dependency.
\end{itemize}

All our experiments were conducted on a Thinkpad P43s with an Intel Core i7 processor and
32GB RAM, running Arch linux, using the same version of \nodejs (14.3.0), to avoid any updates to the runtime 
environment that could affect run times and thus skew the results.

\paragraph{Guarded execution mode.}

For {\bf RQ5}, we repeated the experiments with guards enabled, and measured the running time and size of expanded code for the client test 
suites to determine the increase in overhead due to these extra checks. 

In addition, we report on a case study involving \code{depd}~\citep{depd}, a subject application with a known vulnerability, and on
experiments with {\tt osenv} and {\tt node-os-uptime}, two {\tt npm} modules  with confirmed vulnerabilities that were used as
experimental subjects in \cite{ichneaPaper}%
\footnote{
  Of the subject applications reported on in \cite{ichneaPaper}, these were the only two that had a confirmed vulnerability and
  a test suite with passing tests.
}.

%We invoke \tool on {\it clients} of \code{depd}, finding that the vulnerability is stubbed, and run the client's tests (similarly for transitive clients).
%Our guards report the vulnerability, and this case study is detailed below.
 
\paragraph{Bundlers.} 

For {\bf RQ6}, each subject application was bundled use the \href{https://rollupjs.org/guide/en/}{\textcolor{blue}{\tt{Rollup}}} bundler \citep{rollupNPM}. 
This involved the creation of a bundler configuration file (which we generated automatically given the application's \code{package.json} file)
to bundle the application based on its listed entry points and to create a single bundle that also includes all of its production dependencies%
\footnote{
  The default behavior of \code{rollup} is to ignore dependent modules in \code{node_modules}, but the bundle should all code in which
  stubs may be introduced, to be able to determine \tool's effectiveness. 
}.
We measure and report on the sizes of the resulting bundle, both with and without having applied \tool, to determine what additional size 
reduction is enabled by \tool.

\subsection{\bf Overview of Results}

The results of running \tool on the projects are displayed in Tables \ref{table:SizeRQStaticTable} and \ref{table:SizeRQDynamicTable}.
We show the size of the original source code, the size of the application distribution, and then the resulting size of the distribution after we run our transformation on it, with both the static and dynamic call graphs.

Note that the size immediately after transformation is only representative of the stubbed application size if no stubs are expanded.
To gain a realistic estimate of the size reduction in a standard use-case of the application, we identified five clients for each application and tracked how many stubs were expanded during the execution of the test suites of these clients.
Then, we consider the size of the application to be its base stubbed size \emph{plus} the total size of the stubs that were expanded during the client tests.
This is reported as a range of the lower and upper bounds of application size over the five clients.
The full data is included in the supplementary material. 
% \TODO{need to confirm that this journal allows use to provide supplemental material}

The first row of Table~\ref{table:SizeRQStaticTable} reads: after running \tool with the static call graph, the \code{memfs} source code is reduced the size to 19KB, which is a reduction of 87\% of the original application size.
This expanded to a minimum of 19KB (i.e., nothing was expanded) and a maximum of 138KB over the five clients tested; the expanded code is a reduction of 87\% (with minimum expansion) and 5\% (with maximum expansion) of the original application size.
The first row of Table~\ref{table:SizeRQDynamicTable} can be read the same way, but for results after running \tool with the dynamic call graph on \code{memfs} and testing with the same five clients.

In the remainder of this section, we will address each research question in order.

% \noindent
% We now address each research question in order.

%
%
%
\subsection*{\bf RQ1: How much does \tool reduce application size, and which type of call graphs (static or dynamic) produces smaller applications?}

We refer the reader to Tables~\ref{table:SizeRQStaticTable} and \ref{table:SizeRQDynamicTable}.
In these tables, it can be seen that, using static call graphs, size reductions ranging from 2\% to 87\% are achieved (\avgStubSizeReduction on average). The case where
a size reduction of only 2\% is achieved is \code{commander}, which has no dependencies and appears to be a bit of an outlier. 
Using dynamic call graphs, size reductions ranging from 3\% to 89\% achieved (\avgStubSizeReductionDynamic on average). 
Overall, the use of static call graphs results in larger size reductions in 11/15 cases, and in larger size reductions on average
(\avgStubSizeReduction on average when static call graphs are used vs. \avgStubSizeReductionDynamic when dynamic call graphs are used).

Many of these packages have millions of weekly downloads, and so the size savings add up quickly: for example, \code{css-loader} 
is 2.764MB, and with 10 million weekly downloads we have nearly 28TB of data transferred to users every week.
\tool reduces \code{css-loader}'s initial size by 80\% with both call graphs, which would contribute to 22 fewer TB being 
transferred weekly (for one project!).

\begin{takeaway} 
  On average, \tool reduces initial application size by \avgStubSizeReduction when using static call graphs, and by \avgStubSizeReductionDynamic when using dynamic call graphs.
\end{takeaway}

\subsection*{\bf RQ2: How much code is dynamically loaded due to stub expansion?}

\begin{table*}
{
\begin{subtable}{\textwidth}
    %  \vspace{10pt}
     \centering
     \scriptsize
     \begin{tabular}{ l || c | c || c | c | c | c | c } 
  \multicolumn{1}{l||}{} & \multicolumn{2}{c||}{{\bf Client}} & \multicolumn{5}{c}{{\bf Stubbed code: effect of expansions}} \\
  \hline
  {\bf Proj} & \textbf{Client Proj}& \textbf{Time (s)} & \textbf{Time (s)} & \textbf{Slowdown (\%)} & \textbf{Files} & {\bf Fcts} & {\bf Exp (KB)} \\
  \hline\hline
    & \href{https://github.com/jshjohnson/Choices}{\textcolor{blue}{\texttt{Choices}}} & 5.06 & 5.16   & 2\% & 1 & 0 & 20.06 \\
     & \href{https://github.com/4Catalyzer/found}{\textcolor{blue}{\texttt{found}}} & 30.61 & 31.83  & 4\% & 1 & 0 & 20.06  \\
   {\tt redux} & \href{https://github.com/GriddleGriddle/Griddle}{\textcolor{blue}{\texttt{Griddle}}} & 8.93 & 8.91  & 0\% & 1 & 0 & 20.06 \\
     & \href{https://github.com/atlassian/react-beautiful-dnd}{\textcolor{blue}{\texttt{react-beautiful-dnd}}} & 61.70 & 63.49 & 3\% & 2 & 0 & 20.06  \\
     & \href{https://github.com/omnidan/redux-ignore} {\textcolor{blue}{\texttt{redux-ignore}}} & 0.57 & 0.58 & 2\% & 1 & 0 & 20.06  \\
  \hline\hline
    & \href{https://github.com/bower/decompress-zip}{\textcolor{blue}{\texttt{decompress-zip}}} & 0.70 & 0.74 & 6\% & 1 & 0 & 63.25  \\
    & \href{https://github.com/downshift-js/downshift}{\textcolor{blue}{\texttt{downshift}}} & 1.43 & 1.44 & 1\% & 1 & 0 & 63.25  \\
    {\tt q} & \href{https://github.com/danielzzz/node-ping}{\textcolor{blue}{\texttt{node-ping}}} & 3.80 & 4.20 & 10\% & 1 & 0 & 63.25  \\
    & \href{https://github.com/node-saml/passport-saml}{\textcolor{blue}{\texttt{passport-saml}}} & 0.41 & 0.44 & 6\% & 0 & 0 & 0.00  \\
    & \href{https://github.com/ranm8/requestify}{\textcolor{blue}{\texttt{requestify}}} & 2.92 & 2.99 & 2\% & 1 & 0 & 63.25  \\
  \hline\hline
    & \href{https://github.com/appium/appium-base-driver}{\textcolor{blue}{\texttt{appium-base-driver}}}  & 8.66 & 10.04 & 14\% & 39 & 0 & 146.10 \\
    {\tt body} & \href{https://github.com/expressjs/express}{\textcolor{blue}{\texttt{express}}}  & 1.05 & 1.89 & 45\% & 48 & 0 & 231.69 \\
   {\tt -} & \href{https://github.com/karma-runner/karma}{\textcolor{blue}{\texttt{karma}}}   & 2.08 & 2.12 & 2\% & 40 & 0 & 199.57 \\
    {\tt parser} & \href{https://github.com/moleculerjs/moleculer-web}{\textcolor{blue}{\texttt{moleculer-web}}}  & 5.80 & 6.46 & 10\% & 48 & 0 & 231.69 \\
    & \href{https://github.com/thiagobustamante/typescript-rest}{\textcolor{blue}{\texttt{typescript-rest}}} & 13.17 & 14.89 & 12\% & 48 & 0 & 231.69 \\
  \end{tabular}
   \vspace*{1mm}
  \caption{Stubbed with static call graph}\label{table:clientSpecTableStatic}
  \end{subtable}
  
  \begin{subtable}{\textwidth}
    %  \vspace{10pt}
     \centering
     \scriptsize
     \begin{tabular}{ l || c | c || c | c | c | c | c } 
  \multicolumn{1}{l||}{} & \multicolumn{2}{c||}{{\bf Client}} & \multicolumn{5}{c}{{\bf Stubbed code: effect of expansions}}  \\
  \hline
   {\bf Proj} & \textbf{Client Proj}& \textbf{Time (s)} & \textbf{Time (s)} & \textbf{Slowdown (\%)} & \textbf{Files} & {\bf Fcts} & {\bf Exp (KB)} \\
  \hline\hline
    & \href{https://github.com/jshjohnson/Choices}{\textcolor{blue}{\texttt{Choices}}} & 5.06 & 5.05 & 0\%  & 1 & 0 & 20.06\\
     & \href{https://github.com/4Catalyzer/found}{\textcolor{blue}{\texttt{found}}} & 30.61 & 31.34 & 2\% & 1 & 0 & 20.06\\
   {\tt redux} & \href{https://github.com/GriddleGriddle/Griddle}{\textcolor{blue}{\texttt{Griddle}}} & 8.93 & 9.03 & 1\% & 1 & 0 & 20.06\\
     & \href{https://github.com/atlassian/react-beautiful-dnd}{\textcolor{blue}{\texttt{react-beautiful-dnd}}} & 61.70 & 62.12 & 1\% & 2 & 0 & 20.06\\
     & \href{https://github.com/omnidan/redux-ignore} {\textcolor{blue}{\texttt{redux-ignore}}} & 0.57 & 0.59 & 3\% & 1 & 0 & 20.06\\
  \hline\hline
    & \href{https://github.com/bower/decompress-zip}{\textcolor{blue}{\texttt{decompress-zip}}} & 0.70 & 0.78 & 10\% & 0 & 5 &  2.98 \\
    & \href{https://github.com/downshift-js/downshift}{\textcolor{blue}{\texttt{downshift}}} & 1.43 & 1.44 & 1\% & 0 & 1 &  0.88 \\
    {\tt q} & \href{https://github.com/danielzzz/node-ping}{\textcolor{blue}{\texttt{node-ping}}} & 3.80 & 4.08 & 7\% & 0 & 6 &  2.86 \\
    & \href{https://github.com/node-saml/passport-saml}{\textcolor{blue}{\texttt{passport-saml}}} & 0.41 & 0.42 & 2\% & 0 & 0 &  0.00 \\
    & \href{https://github.com/ranm8/requestify}{\textcolor{blue}{\texttt{requestify}}} & 2.92 & 3.05 & 4\% & 0 & 2 &  0.86 \\
  \hline\hline
    & \href{https://github.com/appium/appium-base-driver}{\textcolor{blue}{\texttt{appium-base-driver}}}  & 8.66 & 9.26 & 6\% & 8 & 0 & 6.69\\
    {\tt body} & \href{https://github.com/expressjs/express}{\textcolor{blue}{\texttt{express}}}  & 1.05 & 1.21 & 14\% & 14 & 0 & 79.70\\
   {\tt -} & \href{https://github.com/karma-runner/karma}{\textcolor{blue}{\texttt{karma}}}   & 2.08 & 2.09 & 1\% & 12 & 0 & 79.03\\
    {\tt parser} & \href{https://github.com/moleculerjs/moleculer-web}{\textcolor{blue}{\texttt{moleculer-web}}}  & 5.80 & 6.38 & 9\% & 14 & 0 & 79.70\\
    & \href{https://github.com/thiagobustamante/typescript-rest}{\textcolor{blue}{\texttt{typescript-rest}}} & 13.17 & 14.48 & 9\% & 14 & 0 & 79.70\\
  \end{tabular}
   \vspace*{1mm}
  \caption{Stubbed with dynamic call graph}\label{table:clientSpecTableDynamic}
  \end{subtable}
  \caption{Results for Clients of Select Projects}
}\end{table*}

Again referring to Table~\ref{table:SizeRQStaticTable} and \ref{table:SizeRQDynamicTable}, this time to the {\bf Expanded KB} range columns, we see that the top end of the expanded ranges using the static call graph are smaller than (or equal to) the expanded ranges using the dynamic call graph in 11/15 cases.
This aligns with our findings in {\bf RQ1}.
In all but one case, the minimal expanded size is close to the reduced application size, and the maximum size increase is  $>2$x in only two cases.

The case where {\tt glob} is processed using a dynamic call graph is an interesting outlier, as its size is {\it larger} than the original code after all stubs have been expanded. 
This is because not much of {\tt glob} is stubbed (the initial size reduction is only 4\%, or 2KB), and the code required to support stub expansion is larger than the initial size 
reduction due to the extra boilerplate that was introduced by \tool (import statements, \code{eval} call, reassignments to imports, etc.).

To break down the results further, we consider the results for all clients of a few packages.
Tables~\ref{table:clientSpecTableStatic} and \ref{table:clientSpecTableDynamic} display all the metrics tracked for all clients of \code{redux}, \code{q}, and \code{body-parser}.
These metrics are the test suite runtimes, the percentage slowdown due running the stubbed code, and number and size of stubs dynamically expanded during the tests.
We chose these applications to display as we felt they are a representative sample of our results; the full data for all clients of all projects is included in the supplementary material.

The first row of Table~\ref{table:clientSpecTableStatic} reads: for \code{redux}, its client application \code{Choices} has an average test suite runtime of 5.06 seconds. 
When the \code{Choices} test suite is rerun with stubbed \code{redux} (via the static call graph), it has an average runtime of 5.16 seconds, which is a slowdown of 2\%; 1 file stub and no function stubs were expanded, and the total size of stubs expanded was 20.06KB.
% When the \code{Choices} test suite is rerun with its \code{redux} dependency replaced with the stubbed \code{redux} (via the static callgraph), it has an average runtime of 5.16 seconds, which is a slowdown of 2\%; 1 file stub and no function stubs were expanded, and the total size of stubs expanded was 20.06KB.
The first row of Table~\ref{table:clientSpecTableDynamic} shows the results of rerunning again with stubbed \code{redux} via the dynamic call graph: now, \code{Choices}' test suite has an average runtime of 5.05 seconds, which is a slowdown of 0\%; 1 file stub and no function stubs were expanded, and the total size of stubs expanded was 20.06KB.

Digging into the client-specific data reveals some interesting trends.
There appears to be a correlation between the number of stubs expanded for the static and dynamic call graphs.
For example, consider the clients of \code{body-parser}: even though there are more stub expansions using the static call graph vs. using the dynamic call graph, it appears that there are ``sets'' of functionality that are commonly expanded together (seen here as whenever 48 file stubs are expanded in the static case, 14 file stubs are expanded in the dynamic case).
The range of expansions among clients suggest that some of the clients use more of an application's untested functionality than others.

We also noted consistency in \emph{which} stubs are expanded.
For example, in the ``sets'' of expanded functionality described earlier, these are the {\it same} 48 and 14 files every time.
As an additional example, all the clients of \code{redux} expand one file stub (one client expands two)---this is always the \emph{same} stub that is expanded.
In the other applications, there is always significant overlap in which stubs are expanded with different clients.
This suggests that some of these applications have commonly used functionalities that are untested, so developers could use this information to shore up their test suites.

% \AT{The \textsc{Doloto} Connection}
% As it happens, \textsc{Doloto}~\citep{DBLP:conf/sigsoft/LivshitsK08} loads functions in clusters (which are determined ahead of time via their ``access profiles''). 
% We can only speculate that the sets of functionality we refer to here might match these clusters, at least in spirit, though we cannot test this claim as \textsc{Doloto} is unable to run on these modern JavaScript applications.

% \AT{Is this something we want to say?}
Finally, we observe that the dynamic call graph typically produces far fewer file stub expansions than the static call graph.
There are a few dimensions to this.
On one hand, as JavaScript is a dynamic language, the static call graph is likely to be incomplete---functions in JavaScript are often called in highly dynamic ways, and these kinds of calls are more easily detected using dynamic analyses.
On the other hand, the dynamic call graph is more susceptible to lower-quality tests: if the application is poorly tested, the dynamic call graph will report many unreachable functions and files.
It is not immediately clear which call graph yields ``better'' results, as fewer stubs mean less size reduction, but also less overhead---we ultimately leave the decision up to the developer.

\begin{takeaway}
  Most package clients load very little code dynamically.
  Many applications have commonly loaded ``sets'' of code, representing broadly used, untested functionality.
\end{takeaway}

\subsection*{\bf RQ3: How much overhead is incurred due to stub expansion?}

To determine the performance overhead introduced by stub expansion, we measured the running times of the test suites of clients of 
applications processed by \tool.

We decided not to aggregate runtime information over all clients of a package as the overhead depends on many factors outside of our control: the number of tests, the structure of the tests, the raw running time of the test, etc.
Instead, we conducted a case study on the effect of the dynamic code loading for the individual clients of the three projects presented in Tables~\ref{table:clientSpecTableStatic} and \ref{table:clientSpecTableDynamic}.
The results for all test applications are included in the supplementary material, but the trends are upheld across the full data. 
% \TODO{make sure that we can have supplemental materials}

Referring to the time columns of Tables~\ref{table:clientSpecTableStatic} and \ref{table:clientSpecTableDynamic}, the following
conclusions can be drawn.
First, a correlation between the slowdown and the number of stub expansions can be observed: as more code is dynamically loaded, the performance overhead increases.
This aligns with our expectations, as stub expansions involve additional I/O and compute time.
That said, the runtime overhead is never extreme, and the slowdowns still leave the running times of the test suites well within the same order of magnitude.
As a percentage, some runtime overhead is high (e.g., \code{body-parser}'s \code{express} dependency), but the magnitude of the change is not (only 0.84 seconds).
We do not see high percentage slowdowns for long-running tests, for instance \code{redux}'s \code{found} and \code{react-beautiful-ignore} clients have 4\% and 3\% slowdowns respectively.
We conjecture that the amount of overhead mostly has to do with the I/O required to load the dynamic code. % , on the order of milliseconds per file (depending on the size of the files being read).

\begin{takeaway} 
By and large, the magnitude and percentage overhead introduced by dynamic loading is small.
\end{takeaway}

\subsection*{\bf RQ4: How much time does \tool need to transform applications?}

Table~\ref{table:TimingResultsTable} shows the time needed by \tool to process each of the \numProjects projects. Here, we distinguish
between the time needed to construct call graphs, and the time needed to transform the source code. 
% \TODO{add "total" columns to the table}

\begin{table}
\centering
{\small
  \begin{tabular}{ l || c | c || c | c }
  \multicolumn{1}{c||}{} & \multicolumn{2}{c||}{{\bf Static CG}} & \multicolumn{2}{c}{{\bf Dynamic CG}} \\
  \hline
  \textbf{Project} & \textbf{CG generation (s)} & {\bf Transf. (s)} & \textbf{CG generation (s)} & {\bf Transf. (s)}\\
  \hline\hline
   \href{https://github.com/streamich/memfs}{\textcolor{blue}{\texttt{memfs}}} &  740.18 & 2.46 & 15.97 & 2.72\\
   \href{https://github.com/bdistin/fs-nextra}{\textcolor{blue}{\texttt{fs-nextra}}} & 380.97 & 1.11 & 13.94 & 1.10\\
   \href{https://github.com/expressjs/body-parser}{\textcolor{blue}{\texttt{body-parser}}} & 295.38 & 3.43 & 10.53 & 3.79 \\
   \href{https://github.com/tj/commander.js}{\textcolor{blue}{\texttt{commander}}} & 554.93 & 1.94 & 24.56 & 1.61 \\
   \href{https://github.com/webpack/memory-fs}{\textcolor{blue}{\texttt{memory-fs}}} & 324.06 & 1.73 & 5.33 & 1.75 \\
   \href{https://github.com/isaacs/node-glob}{\textcolor{blue}{\texttt{glob}}} &  300.80 & 2.09 & 18.66 & 1.46\\
   \href{https://github.com/reduxjs/redux}{\textcolor{blue}{\texttt{redux}}} & 1349.09 & 3.18 & 182.02 & 4.02 \\
   \href{https://github.com/webpack-contrib/css-loader}{\textcolor{blue}{\texttt{css-loader}}} & 1137.77 & 14.85 & 48.61 & 15.52 \\
   \href{https://github.com/kriskowal/q}{\textcolor{blue}{\texttt{q}}} &  336.31 & 4.41 & 10.98 & 4.85\\
   \href{https://github.com/pillarjs/send}{\textcolor{blue}{\tt send}} & 279.16 & 1.75 & 7.57 & 1.67\\
   \href{https://github.com/expressjs/serve-favicon}{\textcolor{blue}{\tt serve-favicon}} & 259.06 & 0.76 & 3.91 & 0.79\\
   \href{https://github.com/expressjs/morgan}{\textcolor{blue}{\tt morgan}} & 313.76 & 1.20 & 8.65 & 1.16\\
   \href{https://github.com/expressjs/serve-static}{\textcolor{blue}{\tt serve-static}} & 276.89 & 1.67 & 7.51 & 1.60\\
   \href{https://github.com/facebook/prop-types}{\textcolor{blue}{\tt prop-types}} &752.94 & 2.12 & 12.79 & 1.89 \\
   \href{https://github.com/expressjs/compression}{\textcolor{blue}{\tt compression}} & 279.18 & 1.28 & 6.78 & 1.37\\
  \end{tabular}
   \vspace*{1mm}
  \caption{Callgraph generation and transformation timing}\label{table:TimingResultsTable}
}\end{table}

The first row of the table reads: for \code{memfs}, generating the static call graph takes 740.18 seconds and applying transformations
based on this call graph takes 2.46 seconds. Furthermore,  generating the dynamic call graph takes 15.97 seconds and applying transformations
based on this call graph takes 2.72 seconds.

From the table, it can be seen that the cost of the code transformation itself is negligible.
The longest runtime is 15 seconds on the \code{css-loader} project, which is unsurprising given that \code{css-loader} is the largest 
subject application (2.76MB).
There is no difference between the transformation times using the static vs dynamic call graphs.
This is also unsurprising, as the same process is used to run the transformation in either case, and, generally, a similar number of stubs is created.
In cases such as \code{q}, where the dynamic call graph produces a larger stubbed application and yet it takes longer to run, this is because there are more 
\emph{function} stubs being generated (compared to a single file stub being generated when using static call graphs).

The cost of call graph construction is more noteworthy.
Overall, we see that constructing a static call graph takes one to two orders of magnitude more time than constructing the dynamic call graph.
We also observe a correlation between the times to construct the static and dynamic call graphs.
To construct the dynamic call graph, \tool simply computes a coverage report from running an application's tests (including \code{node_modules}), which amounts to the time to run the tests plus some small overhead.
The slower runtime of static call graph construction is due to our inclusion of the generation of the CodeQL database in the overall runtime, which is directly proportional to the amount of code in the project 
(in order to run any static analysis queries, CodeQL must build a database of the application's code---this is a one-time cost as long as the code does not change).
% and database generation time is directly proportional to the amount of code in the project.

We envision the use-case of \tool to be a final stage in the creation of a production release, and so we do not believe that a build-time of 5-15 minutes to be prohibitive.
If time is an issue or if a user wanted to apply \tool more frequently, they could opt for using dynamic call graphs.

\begin{takeaway} 
  The average runtime of \tool with the static call graph is not prohibitive (at roughly 8.3 minutes), and is much lower (28 seconds) with the dynamic call graph.
\end{takeaway}

\subsection*{\bf RQ5: How much run-time overhead is incurred by guarded execution mode and can it detect security vulnerabilities?}\label{sec:evalRQ}

The use of ``dangerous'' functions such as \code{eval} and \code{exec} that interpret string values as code is known to cause injection
vulnerabilities in JavaScript applications \cite{ichneaPaper}.
It is particularly concerning if such functions are invoked from untested code, because it means that the developers may not have considered
all situations where calls to such functions are executed.  
\tool's guarded execution mode aims to mitigate this risk, by adding dynamic checks for such functions in stubbed-out code so that a warning
can be issued or execution can be terminated when such calls are encountered. 
These dynamic checks may have a noticeable impact on code size and execution times, and research question RQ5 aims to 
establish the magnitude of that effect.

We first consider performance and code size by repeating the experiments in guarded execution mode.
The initial distribution sizes for the 15 applications is the same, but we noted an increase in expanded code sizes, which in many
cases now exceeds the size of the original application. This is unsurprising, as the code size overhead of the guards is significant.
Consider Tables~\ref{table:clientSpecTableEvalStatic} and \ref{table:clientSpecTableEvalDynamic}%
\footnote{
  The full data for all applications is included in the supplemental material. 
}, which report experimental data for the 
{\tt q} package's five clients. The first row of Table \ref{table:clientSpecTableEvalStatic} reads: for the \code{decrompress-zip} client 
of \code{q}, the test suite runs in 1.22s which is a slowdown of 43\% over running the test suite with the original \code{q} package.
% \TODO{check that the previous sentence is correct}
Moreover,  during these tests 240.9KB of code is expanded, as compared to 63.25KB of code being expanded without guarded execution mode 
(this last column is also included in Table \ref{table:clientSpecTableStatic}).
Note that, when using static call graphs, the expanded code size is almost 4x larger when guards are enabled.
The performance of the code also degrades, though the raw numbers are again fairly low---we again suspect the increased slowdown to be (mostly) 
due to fact that the program needs to load more code.
That said, we did observe some significant overhead in longer-running applications, for instance slowdowns of 19\% and 3\% in 
the longer running test suites of \code{redux}'s \code{found} and \code{react-beautiful-dnd} clients, respectively (when using static call graphs), 
as compared with 4\% and 3\% respectively without guarded execution mode.

\begin{table}
{\small
  \begin{subtable}{\textwidth}
  \centering
  \begin{tabular}{ l || c | c | c || c } 
  \multicolumn{1}{c}{} & \multicolumn{3}{c}{{\bf With guards}} \\
  \textbf{Client Proj} & \textbf{Time (s)} & \textbf{Slowdown (\%)} & {\bf Exp. KB} & {\bf Exp. KB no guards} \\
  \hline\hline
    \href{https://github.com/bower/decompress-zip}{\textcolor{blue}{{\texttt{decompress-zip}}}} & 1.22 & 43\% & 240.9 & 63.3 \\
    \href{https://github.com/downshift-js/downshift}{\textcolor{blue}{{\texttt{downshift}}}} & 1.47 & 3\% & 240.9 & 63.3 \\
    \href{https://github.com/danielzzz/node-ping}{\textcolor{blue}{{\texttt{node-ping}}}} & 4.89 & 22\% & 240.9 & 63.3 \\
    \href{https://github.com/node-saml/passport-saml}{\textcolor{blue}{{\texttt{passport-saml}}}} & 0.48 & 13\% & 0.0 &  0.0 \\
    \href{https://github.com/ranm8/requestify}{\textcolor{blue}{{\texttt{requestify}}}} & 3.30 & 12\% & 240.9 & 63.3 \\
  \end{tabular}
   \vspace*{1mm}
   \caption{Stubbed with static call graph}\label{table:clientSpecTableEvalStatic}
  \end{subtable}
  
  \begin{subtable}{\textwidth}
  \centering
  \begin{tabular}{ l || c | c | c || c} 
  \multicolumn{1}{c}{} & \multicolumn{3}{c}{{\bf With guards}} \\
  \textbf{Client Proj}& \textbf{Time (s)} & \textbf{Slowdown (\%)} & {\bf Exp. KB} & {\bf Exp. KB no guards} \\
  \hline\hline
    \href{https://github.com/bower/decompress-zip}{\textcolor{blue}{{\texttt{decompress-zip}}}} & 0.78 & 10\% & 16.6 & 3.0 \\
    \href{https://github.com/downshift-js/downshift}{\textcolor{blue}{{\texttt{downshift}}}} & 1.63 & 13\% & 3.2 & 0.9 \\
    \href{https://github.com/danielzzz/node-ping}{\textcolor{blue}{{\texttt{node-ping}}}} & 4.69 & 19\% & 14.9 & 2.9 \\
    \href{https://github.com/node-saml/passport-saml}{\textcolor{blue}{{\texttt{passport-saml}}}} & 0.51 & 19\% & 0.0 & 0.0 \\
    \href{https://github.com/ranm8/requestify}{\textcolor{blue}{{\texttt{requestify}}}} & 3.43 & 15\% & 4.3 & 0.9 \\
  \end{tabular}
    \caption{Stubbed with dynamic call graph}\label{table:clientSpecTableEvalDynamic}
   \vspace*{1mm}
  \end{subtable}
  \caption{Results for Clients of \code{q} with guards enabled}
}\end{table}

%Overall, we interpret these results to suggest that enabling guards leads to the same reduction in initial application size, but it does increase
%the size after stub expansion noticeably.

\paragraph{Detecting security vulnerabilities.}

%We envision a deployment scenario where guards are enabled during a testing phase during which package developers want to find security 
%vulnerabilities in their applications.  To establish whether or not this is realistic, we report on a case study involving the 
%\code{depd}~\citep{depd} application.

When guarded execution mode was enabled, calls to \code{eval} were intercepted in running the test suites of three subject applications:
   \code{body-parser}, \code{send}, and \code{serve-static}.
Upon investigation, we found that the dangerous calls were not in the code of these packages themselves, but \textit{hidden in one their dependencies}.
Specifically, all these packages rely on an \emph{old version} of \code{depd}~\citep{depd}: \code{body-parser} and \code{send} have a direct dependency, 
and \code{serve-static} has a transitive dependency as it depends on \code{send}.
We confirmed that this is indeed a problem by examining the \code{depd} project repository on Github and found that the
problematic \code{eval} was removed on January 12, 2018 with commit~\citep{depd-removal}, which fixed three issues, \citep{depd-issue1}, \citep{depd-issue2}, and \citep{depd-issue3}.
These issues were filed because \code{eval} is not only bad practice, but its use is disallowed in Chrome apps and Electron apps. %, both of which are very popular platforms (we are paraphrasing points brought up in the linked GitHub issues).
To fix this issue, we removed the lock on the {\tt depd} version (i.e., set it to \code{*}) to get the applications to use the  current version of {\tt depd}, and confirmed that all client tests still pass.

To further test the effectiveness of guarded execution mode, we ran another experiment involving two other applications with known vulnerabilities: \href{https://github.com/npm/osenv}{\color{blue}{\tt osenv}} and  \href{https://github.com/oroce/node-os-uptime}{\color{blue}{\tt node-os-uptime}}.
These projects were used as experimental subjects%
\footnote{
  Of all the subject applications considered in \citep{ichneaPaper}, these are the only two that still build, install, and have a test suite with passing tests, as required by \tool. 
} in the evaluation of a dynamic taint analysis  \citep{ichneaPaper} that detected vulnerabilities in them.
In both projects, a function containing a call to a dangerous function (\code{exec} in the case of \code{osenv} and \code{execSync} in the case of \code{node-os-uptime}) 
was stubbed out by \tool. We created a new test containing the same code fragment that was used in \citep{ichneaPaper} to detect the vulnerability, 
and confirmed that the guard introduced by \tool was triggered when the test was executed.  
% \TODO{I think we should consistently turn all URLs into citations, and not rely on hyperlinks as in the above text}
 
\begin{takeaway} 
  Guarded execution mode allows developers to detect injection vulnerabilities in imported modules of which developers may be unaware, and 
  we found several examples of this in our experiments.
\end{takeaway}

% \subsection{Bundlers}
\subsection*{\bf RQ6: How much does \tool reduce the size of applications that have been bundled using \texttt{Rollup}?}\label{sec:evalRQ}

\begin{table}
\centering
{\small
  \begin{tabular}{ l || c | c || c | c || c | c } 
  \multicolumn{3}{c}{} & \multicolumn{4}{c}{\textbf{Stubbed Bundle}}  \\
  \multicolumn{1}{c||}{} & \multicolumn{2}{c||}{{\bf Bundle}} & \multicolumn{2}{c||}{{\bf Dynamic CG}} & \multicolumn{2}{c}{{\bf Static CG}}  \\
  \hline 
  \textbf{Package} & \textbf{Size (KB)} & \textbf{Red \%} & \textbf{Size (KB)} & \textbf{Red \%} & \textbf{Size (KB)} & \textbf{Red \%} \\
  \hline\hline
   \href{https://github.com/streamich/memfs}{\textcolor{blue}{\texttt{memfs}}} &  128 & 53\% & 10 & 92\% &  10 & 92\%\\
   \href{https://github.com/bdistin/fs-nextra}{\textcolor{blue}{\texttt{fs-nextra}}} & 52 & 0\% & 21 & 60\% &  21 & 60\%\\
   \href{https://github.com/expressjs/body-parser}{\textcolor{blue}{\texttt{body-parser}}} &  626 & 36\% & 534 & 15\% &  534 & 15\%\\
   \href{https://github.com/tj/commander.js}{\textcolor{blue}{\texttt{commander}}} &  72 & 17\% & 47 & 35\% &  47 & 35\%\\
   \href{https://github.com/webpack/memory-fs}{\textcolor{blue}{\texttt{memory-fs}}} &  100 & 17\% & 62 & 38\% &  62 & 38\%\\
   \href{https://github.com/isaacs/node-glob}{\textcolor{blue}{\texttt{glob}}} &  84 & 2\% & 42 & 50\% &  42 & 50\%\\
   \href{https://github.com/reduxjs/redux}{\textcolor{blue}{\texttt{redux}}} &  22 & 92\% & 7 & 67\% &  7 & 67\%\\
   \href{https://github.com/webpack-contrib/css-loader}{\textcolor{blue}{\texttt{css-loader}}} &  962 & 59\% & 393 & 59\% &  393 & 59\%\\
   \href{https://github.com/kriskowal/q}{\textcolor{blue}{\texttt{q}}} &  66 & 77\% & 53 & 19\% &  53 & 19\%\\
   \href{https://github.com/pillarjs/send}{\textcolor{blue}{\tt send}} & 130 & 43\% & 89 & 31\% &  89 & 31\%\\
   \href{https://github.com/expressjs/serve-favicon}{\textcolor{blue}{\tt serve-favicon}} & 18 & 21\% & 12 & 31\% &  12 & 31\%\\
   \href{https://github.com/expressjs/morgan}{\textcolor{blue}{\tt morgan}} & 54 & 3\% & 30 & 44\% &  30 & 45\%\\
   \href{https://github.com/expressjs/serve-static}{\textcolor{blue}{\tt serve-static}} & 107 & 22\% & 95 & 11\% &  95 & 11\%\\
   \href{https://github.com/facebook/prop-types}{\textcolor{blue}{\tt prop-types}} &  {\bf NA} &  {\bf NA} &  {\bf NA} &  {\bf NA} &  {\bf NA} &  {\bf NA}\\
   \href{https://github.com/expressjs/compression}{\textcolor{blue}{\tt compression}} & 23 & 66\% & 21 & 7\% &  21 & 7\%\\
  \end{tabular}
   \vspace*{1mm}
  \caption{Effect of stubbifying bundled projects}\label{table:bundlerExpTable}
}\end{table}

% Earlier in the paper, in Section~\ref{sec:bundler-bg}, we overviewed JavaScript bundlers, tools designed to reduce application size and create smaller distributions.
% Conceptually, \tool should be able to further reduce the size of applications which utilize bundlers, as they are not equipped with a fallback mechanism like dynamic code loading and thus must be certain that the code they remove will never be called.
%Our approach is able to synergize with bundlers to create even smaller distributions.
%To support this, we pose and answer our sixth research question with an experiment wherein we configured each of our subject applications to use the {\tt rollup} bundler, and applied \tool to the resultant bundle.

To answer this research question, we conducted an experiment where we applied the the {\tt Rollup} bundler to each subject application, and applied \tool to
 the resulting bundle. Table \ref{table:bundlerExpTable} displays the results of this experiment.
The first row of this table can be read as follows: for the \code{memfs} project, the size of the rollup bundle is 128KB, which is a reduction of 53\% from the original size of the project. 
When we stubbify that bundle using the dynamic callgraph as input, the result is a bundle of 10KB, which is a further reduction of 92\% from the original bundle.
When we stubbify the bundle instead with the static callgraph as input, the result is also a bundle of 10KB, with the same reduction of 92\% from the original bundle.

Not all of the applications lend themselves well to bundling, and there are a few notable results.
\code{commander} and \code{q} are configured such that when the bundler is applied, the entire package is wrapped in a single function that is called to generate the module exports.
Since this function does not exist in the original module, it is not detected as reachable from the application's tests (since these exercise the original, un-bundled application).
To address this, we configured \tool to prevent it from replacing 4 functions with stubs 
(one in \code{commander}, and three in \code{q}) (recall from Section~\ref{sec:generating-stubs} that programmers can specify in a comment that \tool should not stub a function or file).
Beyond these, \code{prop-types} could not be bundled as it depends on some BabelJS libraries that throw errors when the code format is changed by the bundler, and \code{fs-nextra} is also notable, as it has no dependencies so bundling it does not reduce its size at all.

That said, in every case, we see that \tool achieves additional size reductions on applications after they are bundled, with an average of \avgSizeReductionOnBundles further size reduction.
Indeed, the purpose of bundlers is not to reduce application size, and that is merely a secondary benefit: the main goal of a bundler is to produce  a single file that can be distributed for ease-of-use.
\tool can debloat these bundles, and in every case we find that \tool helps bundlers create even smaller bundles.

To confirm that the debloated bundles behave as expected, we conducted an experiment in which we reconfigured 
the test suites of \code{commander}, \code{body-parser}, and \code{node-glob} to use the debloated bundle%
\footnote{
  In general, adapting application test suites to work with a bundled version of the application instead of the original version 
  can be a complex and error-prone process, as test suites may import specific functions (that may be renamed by the bundler) from specific 
  files (that may be combined by the bundler). For the applications mentioned here, this conversion was straightforward.
}, and found that the project tests executed as expected.

\begin{takeaway} 
  \tool achieves significant code size reductions when applied to bundled applications, by reporting a further size reduction of \avgSizeReductionOnBundles on top of the reduction already afforded by bundlers.
\end{takeaway}

% \TODO{I'm not at all happy with the discussion about the interaction between \tool and bundlers: the last paragraph makes it sound like it
%   is actually counterproductive to run our tool in such cases because it creates a situation where code that is needed is always placed
%   in a stub. We need to discuss if/how we can portray this in a better light. Would it make sense to configure \tool so that it never
%   stubs out functions that are introduced by a bundler?}

\section{Threats to Validity}

Our approach relies on an application's test suite as the entry point for call graph construction.
This entwines the performance of our tool with the quality of the tests.
An application with a low-quality test suite may generate a call graph that does not represent a comprehensive usage of the application functions, thus leading to more stubs and likely more stub expansion.
To mitigate against bias, we did not consider the {\it quality} of an application's tests when selecting projects for our evaluation, only that the application had tests at all (and that these tests passed).
Concretely, Table~\ref{table:projDescTable} shows that applications have differing numbers of tests, as many as 1706 and as few as 30, with every application having over 10K LOC, suggesting that the quality of the test suites of the projects in our evaluation varies. 

Also, we are cognizant that we are drawing generalized conclusions based on a limited set of JavaScript projects.
To mitigate potential bias in project selection, we selected 15 projects in a systematic manner from the most popular projects published by \npm: from a list of projects sorted in descending order by number of weekly downloads, we attempted to install, build, and run project test suites.
If a project satisfied all these criteria, we then randomly selected from its clients and attempted to install, build, and run their tests; if the project had five such clients, it was selected.
We also note that the subject applications vary considerably in size, in number of dependencies, as well as application domains: e.g., {\tt memfs} is an in-memory file system, {\tt body-parser} is a parser for request bodies, and {\tt css-loader} is a custom loader for {\tt css} files.

It is also possible that the reported runtimes are subject to measurement bias.
We mitigate this by running all performance experiments on a machine with no other processes running.
We also report the average run time over 10 runs, after discarding two initial runs, which minimizes risk of long experiment startup time.

In our experiments with the Rollup bundler, we had to manually configure \tool to avoid stubbing four functions in the bundles for 
{\tt commander} and {\tt q} that were introduced by the bundler.  Since these functions did not occur in the call graphs created by 
\tool, they would otherwise have been replaced with stubs, resulting in size reductions in excess of 95\%. However, such a size reduction
would have been counterproductive---these functions are always executed when the bundles are used, and thus the introduced stubs would always have to be expanded.   
%
%% as these functions (a) encompassed huge amounts of application functionality and (b) were not detected as reachable by the analyses.
%%Had we not done this, the reported size reductions would have been in excess of 95\%.
There is a potential for human error here, but identifying these four functions was not difficult:
for {\tt commander}, the bundler wrapped the entire module in an immediately invoked function expression (IIFE), and in the case of {\tt q} 
the bundler included large swaths of code in the exported object of the bundle. Longer term, an automated solution to this problem could be devised.

\section{Related Work}
  \label{sec:RelatedWork}
  
% \TODO{please review the discussion of Doloto -- some text from section 2 was incorporated here} 
% \EA{looks good!}
  
Our work was inspired by  \textsc{Doloto} \citep{DBLP:conf/sigsoft/LivshitsK08}, a tool that applies code-splitting to an application 
based on  ``access profiles'' obtained from users interacting with an instrumented version of the application.
These access profiles define clusters of functions that should be loaded together, and functionality that should be part of the distribution 
of an application. Applications processed by \textsc{Doloto} ship with enough functionality for initialization, and inessential functions are 
replaced with small stubs that are either replaced once their original code is loaded lazily, or on-demand when a stubbed function is invoked.

There are several factors that make \tool more practical than Doloto. Most importantly, \tool is fully automatic, debloating an application 
based on call graphs that were constructed from its tests.  \tool handles modern JavaScript \cite{ECMAScript2019}, which
includes many features (e.g., classes, promises, async/await, generators, modules etc.) that were not present when Doloto was developed in 2008.
Moreover, \tool supports not only the function-level stubs that were used by Doloto, but also file-level stubs to handle the common case where
all functions in a file are found to be unreachable. \tool also provides a guarded execution mode, which prevents injection vulnerabilities 
resulting from  calls to functions such as \code{eval} and \code{exec} when they are invoked from within untested code that resulted from expanding 
stubs. Lastly, \tool has been developed to be used in conjunction with bundlers.

There is also recent work on debloating other languages: JShrink~\citep{jshrinkPaper} is a tool for debloating the bytecode of Java applications.
Their technique makes use of a combination of both static and dynamic analyses, to use both the strong type guarantees of the Java language, and to also deal with dynamic language features that are becoming more prevalent in modern Java use.

Building minimal application bundles is both well-studied and prevalent in industry.
Several implementations of Smalltalk developed in the 1990s (e.g., \citep{VisualWorks:95,VisualAgeSmalltalk:97}) include features for ``packaging'' or ``delivering'' applications, and IBM's 1997 Handbook for VisualAge for Smalltalk \citep{VisualAgeSmalltalk:97} 
describes a reference-following strategy to determine minimal code for a package.
Compacting code is a related area, for example  \citep{DeSutter:2002:SOM:583854.582445} present Squeeze++, a link-time code compactor for low-level C/C++ code. 
Another facet of this area is specializing distributions: \citep{DBLP:conf/kbse/SharifAGZ18} present TRIMMER, which specializes LLVM bytecode applications to their deployment context using input specialization.
The performance impact of using application bundles has also been studied in the context of Java, where \citep{DBLP:conf/jvm/HovemeyerP01} study performance issues that arise when bundles of JVM class files for Java applications are downloaded from a server.
Broadly, these approaches rely on some form of ``application profile'' obtained via program analysis---\tool builds this profile via static or dynamic analysis of application tests.

In a similar vein, trimming optional functionality from applications has been studied by  ~\citep{Bhattacharya:2013:CCI:2509136.2509522}, who propose an approach relying on a combination of human input, dynamic analysis, and static analysis to identify optional functionality.
\citep{KooEtAl:19} present a technique relying on manual analysis of configuration files and profiling to obtain coverage information for executions in different configurations, minimizing based on that coverage.

Much existing work is concerned with entirely removing unused code.
\citep{AgesenUngar:94} present an type-inference based application extractor for Self~\citep{DBLP:conf/ecoop/AgesenPS93} which extracts a bloat-free source file for distribution.
The Jax application extractor for Java \citep{DBLP:conf/oopsla/TipLSS99} relies on efficient type-based call graph construction algorithms such as RTA \citep{DBLP:conf/oopsla/BaconS96} and XTA \citep{DBLP:conf/oopsla/TipP00} to detect unreachable methods, and further relies on a specification language \citep{DBLP:conf/sigsoft/SweeneyT00} in which users specify classes and methods that are accessed reflectively, going above-and-beyond dead code elimination with, e.g., class hierarchy compaction~\citep{DBLP:journals/toplas/TipSLES02}.
% Beyond detecting and removing unused code, Jax performs additional size-reducing transformations such as method inlining, renaming of program entities, and compaction of the class hierarchy \citep{DBLP:journals/toplas/TipSLES02}, and can create self-contained application bundles.
Rayside and Kontogiannis \citep{DBLP:journals/scp/RaysideK02} present a tool for extracting subsets of Java libraries using Class Hierarchy Analysis \citep{DBLP:conf/ecoop/DeanGC95} to identify the subset of a library that is required by a specific application, though their work does not consider unsoundness.  

On the other hand, the use of code splitting techniques has been explored previously in different contexts.
Besides \textsc{Doloto}~\citep{DBLP:conf/sigsoft/LivshitsK08}, \citep{DBLP:conf/oopsla/KrintzCH99} proposed a code-splitting technique for Java that partitions classes into separate ``hot'' and ``cold'' classes to avoid transferring code that is rarely used.
\citep{DBLP:journals/scp/WagnerGF11} present an optimistic compaction technique for Java applications, where minimized distributions are outfitted with a custom class loader that performs partial loading and on-demand code addition.

% Techniques besides these have also been studied, for instance Heo et al. \citep{DBLP:conf/ccs/HeoLPN18} present an automated technique using delta debugging \citep{DBLP:conf/esec/Zeller99} and model-based reinforcement learning.
% Quach et al. \citep{DBLP:conf/uss/QuachPY18} present an approach where functionality-specific meta-information is embedded into code modules during compilation, so that when a symbol is loaded at run time, the meta-information is consulted so that only the required functionality is loaded. 

%
%
%
%
\subsection{Control Flow Integrity}

The guarded execution mode resembles works on Control Flow Integrity (CFI) verification by, e.g., \citep{abadiccs05}. 
A CFI policy dictates that program execution must follow a predetermined path of a control flow graph, enforced via program rewriting and runtime monitoring.
Conceptually, our guarded execution mode enforces a policy where program execution cannot invoke a predefined list of functions.
\citep{zhangsc13} present a CFI approach that enforces a policy preventing jumps to any but a white-list of locations, whereas our guarded mode enforces a black-list of functions.
% They calculate the white-list via an imprecise but less costly CFG builder tracking only function pointers and return addresses, rather than a very precise CFG builder.
\citep{niuccs15} develop a ``per-input'' CFI technique to avoid the overhead of constructing a control flow graph, and our mode avoids this altogether by pre-transforming code to intercept calls.

\section{Conclusion}

JavaScript is an increasingly popular language for server-side development, thanks in part to the Node.js runtime environment and the vast ecosystem of modules available on {\tt npm}.
Unfortunately, {\tt npm} installs modules with {\it all} of their functionality, even if only a fraction is needed, which causes an undue increase in code size.
In this paper, we presented a fully automatic technique that identifies dead code by constructing static or dynamic call graphs from the application's tests, and replaces code deemed unreachable with either file- or function-level stubs that can fetch and execute the original code dynamically.
The technique also gives users the option to guard their applications against injection vulnerabilities in untested code that result from stub expansion.
This technique is implemented in a tool called \tool, which supports the ECMA\-Script 2019 standard.

In an empirical evaluation on 15 Node.js applications and \numTotalClients clients of these applications, \tool reduced application size by \avgStubSizeReduction on average while incurring only minor performance overhead.
The evaluation also showed that \tool's guarded execution mode is capable of preventing several known injection vulnerabilities that are manifested in stubbed-out code.
Finally, \tool works alongside bundlers, and for the subject applications under consideration, 
we measured an average size reduction of \avgSizeReductionOnBundles in distributions produced by bundlers.

%\begin{acknowledgements}
%\end{acknowledgements}


\bibliographystyle{spbasic}      % basic style, author-year citations
\bibliography{paper}   % name your BibTeX data base

\end{document}